\documentclass[12pt,english,floatfix,nofootinbib,superscriptaddress,aps,prd,preprint]{revtex4}
\usepackage[utf8]{inputenc}
\usepackage{float}
\usepackage{array}
\usepackage{lipsum}
\usepackage{dsfont}
\usepackage{commath}
\usepackage{graphicx}
\usepackage{amsmath,amsthm,amsfonts,amssymb}
\usepackage{graphicx}
\usepackage[english]{babel} 
\usepackage{color}
\usepackage{tensor}
\usepackage{esint}
\usepackage[dvips]{epsfig}
\usepackage[dvips]{graphicx}
\usepackage{float}
\usepackage{units}
\usepackage{textcomp}
\usepackage{mathrsfs}
\usepackage{amsmath}
\usepackage[makeroom]{cancel}
\usepackage{amssymb}
\usepackage{amsbsy}
\usepackage{amsfonts}
\usepackage{amssymb,mathrsfs,xcolor}
\usepackage{esint}
\usepackage{braket}
\usepackage{array}
\usepackage{graphicx}
\usepackage{caption}
\usepackage{subcaption}
\usepackage{comment}

\usepackage{wasysym}
\usepackage{multirow}
\usepackage{wrapfig}

\usepackage{stmaryrd}
\usepackage{upgreek}

\makeatletter

\makeatletter\usepackage{babel}

\usepackage{hyperref}
\hypersetup{
    colorlinks,
    citecolor=blue,
    filecolor=green,
    linkcolor=purple,
    urlcolor=red,
}

\usepackage{slashed}

\newcommand{\ie}{\begin{equation}}
\newcommand{\fe}{\end{equation}}
\newcommand{\se}{\begin{eqnarray}}
\newcommand{\ff}{\end{eqnarray}}

\begin{document}

\title{Gravitational signatures of a non--commutative stable black hole}


\author{N. Heidari}
\email{heidari.n@gmail.com}

\affiliation{Faculty of Physics, Shahrood University of Technology, Shahrood, Iran}

\author{H. Hassanabadi}
\email{hha1349@gmail.com}

\affiliation{Faculty of Physics, Shahrood University of Technology, Shahrood, Iran}
 \affiliation{Department of Physics, University of Hradec Kr$\acute{a}$lov$\acute{e}$, Rokitansk$\acute{e}$ho 62, 500 03 Hradec Kr$\acute{a}$lov$\acute{e}$, Czechia.}

\author{A. A. Araújo Filho}
\email{dilto@fisica.ufc.br (The corresponding author)}

\affiliation{Departamento de Física Teórica and IFIC, Centro Mixto Universidad de Valencia–CSIC, Universidad de Valencia, Burjassot, 46100, Valencia, Spain}

\affiliation{Departamento de Física, Universidade Federal da Paraíba, Caixa Postal 5008, 58051-970, João Pessoa, Paraíba,  Brazil.}

\author{J. K\u{r}\'\i\u{z}}
\email{jan.kriz@uhk.cz}

\affiliation{Department of Physics, University of Hradec Kr$\acute{a}$lov$\acute{e}$, Rokitansk$\acute{e}$ho 62, 500 03 Hradec Kr$\acute{a}$lov$\acute{e}$, Czechia.}

\author{S. Zare}
\email{soroushzrg@gmail.com}
\affiliation{Department of Physics, University of Hradec Kr$\acute{a}$lov$\acute{e}$, Rokitansk$\acute{e}$ho 62, 500 03 Hradec Kr$\acute{a}$lov$\acute{e}$, Czechia.}

\author{P. J. Porfírio}
\email{pporfirio@fisica.ufpb.br}

\affiliation{Departamento de Física, Universidade Federal da Paraíba, Caixa Postal 5008, 58051-970, João Pessoa, Paraíba,  Brazil.}


\date{\today}

\begin{abstract}

This work investigates several key aspects of a non--commutative theory with mass deformation. We calculate thermodynamic properties of the system and compare our results with recent literature. We examine the \textit{quasinormal} modes of massless scalar perturbations using two approaches: the WKB approximation and the Pöschl--Teller fitting method. Our results indicate that stronger non--commutative parameters lead to slower damping oscillations of gravitational waves and higher partial absorption cross sections. Furthermore, we study the geodesics of massless and massive particles, highlighting that the non--commutative parameter $\Theta$ significantly impacts the paths of light and event horizons. Also, we calculate the shadows, which show that larger values of $\Theta$ correspond to larger shadow radii, and provide some constraints on $\Theta$ applying the observation of Sgr $A^{*}$ from the Event Horizon Telescope.
Finally, we explore the deflection angle in this context.


\end{abstract}

\maketitle



\section{Introduction}

General relativity (GR) is a geometric theory of gravity that exhibits intrinsically nonlinear behavior. As a result, exact solutions to its field equations can be challenging to solve, even when additional symmetries and restrictions are taken into account \cite{wald2010general, misner1973gravitation}. A common approach is encountered in the literature in order to overcome this issue: the weak field approximation. This method greatly simplifies the field equations, making them more manageable to work with, and gives rise to gravitational waves, which are of significant interest for studying stability, Hawking radiation of black holes (BHs), and the interactions of BHs with their surroundings in astrophysical scenarios.

Understanding gravitational waves and their associated characteristics is crucial for studying a wide range of physical processes, including cosmological events that took place in the early universe and astrophysical phenomena such as the evolution of stellar oscillations \cite{unno1979nonradial, kjeldsen1994amplitudes, dziembowski1992effects} and binary systems \cite{pretorius2005evolution, hurley2002evolution, yakut2005evolution, heuvel2011compact}. These waves exhibit a range of intensities and characteristic modes, and their spectral properties depend on the generating phenomenon \cite{riles2017recent}. Also, the emission of gravitational waves from BHs is a prominent aspect worthy to be investigated. When a BH is formed through the gravitational collapse of matter, it emanates radiation that includes a bundle of characteristic frequencies unrelated to the process which generated it. These perturbations are called the \textit{quasinormal} modes. More so, the weak field approximation has been widely used in the literature to investigate the \textit{quasinormal} modes of BHs in GR \cite{rincon2020greybody, santos2016quasinormal, oliveira2019quasinormal, berti2009quasinormal, horowitz2000quasinormal, nollert1999quasinormal, ferrari1984new, kokkotas1999quasi, london2014modeling, maggiore2008physical, flachi2013quasinormal, ovgun2018quasinormal, blazquez2018scalar, roy2020revisiting, konoplya2011quasinormal}, Ricci--based gravity theories \cite{kim2018quasi, lee2020quasi, jawad2020quasinormal}, Lorentz violation \cite{maluf2013matter, maluf2014einstein}, and other related fields \cite{JCAP1, JCAP2, JCAP3, jcap4, jcap5}.

Significant progress has been made in the development of gravitational wave detectors, which have enabled the detection of gravitational waves emitted from various physical processes \cite{abbott2016ligo, abbott2017gravitational, abbott2017gw170817, abbott2017multi}. Ground--based interferometers such as VIRGO, LIGO, TAMA-300, and EO-600 have been used to achieve this milestone \cite{fafone2015advanced,abramovici1992ligo,coccia1995gravitational,luck1997geo600}. These detector accuracies have improved over time, bringing them closer to the true sensitivity scheme \cite{evans2014gravitational}. In addition, they have brought about valuable information on the composition of astrophysical objects such as boson and neutron stars. One of the most remarkable aspects of these processes is their connection to BH physics. The gravitational radiation emitted by a perturbed BH carries a signature that can directly verify its existence \cite{thorne2000probing}. Furthermore, the stability of Schwarzschild BHs was initially studied by Regge and Wheeler, followed by Zerilli \cite{regge1957stability,zerilli1970effective,zerilli1974perturbation}.

In recent years, the study of gravitational solutions involving scalar fields has gained much attention due to their intriguing and distinctive characteristics, leading to many astrophysical applications. Among them, BHs with nontrivial scalar fields appear to contradict the well--known no--hair theorem \cite{herdeiro2015asymptotically}, long--lived scalar field patterns \cite{ayon2016analytic}, boson stars \cite{colpi1986boson,palenzuela2017gravitational,cunha2017lensing}, and exotic astrophysical scenarios in the form of gravastars \cite{visser2004stable,pani2009gravitational,chirenti2016did}. Additionally, many feasible phenomena can be derived by considering the Klein--Gordon scalar fields on curved backgrounds, such as BH bombs \cite{cardoso2004black,sanchis2016explosion,hod2016charged} and superradiance \cite{brito2015black}.

The formalism used to describe the geometry of spacetime in GR lacks a bound on the precision of distance measurements, which is believed to be given by the Planck length. To address this issue, one common approach is using the concept of non--commutative (NC) spacetimes. NC geometry is motivated by string/M--theory \cite{szabo2006symmetry,szabo2003quantum,3}, and it has important applications in supersymmetric Yang--Mills theories within the superfield formalism \cite{ferrari2003finiteness,ferrari2004superfield,ferrari2004towards}. Moreover, the Seiberg--Witten map is typically used to introduce non--commutativity in the context of gravity by gauging an appropriate group \cite{chamseddine2001deforming}. In the NC framework, significant progress has been made in the study of BHs \cite{nicolini2009noncommutative,lopez2006towards,modesto2010charged,mann2011cosmological,1,2,campos2022quasinormal,zhao2023quasinormal,karimabadi2020non}, including their evaporating aspects \cite{myung2007thermodynamics,araujo2023thermodynamics} and thermodynamic properties \cite{banerjee2008noncommutative,lopez2006towards,sharif2011thermodynamics,nozari2006reissner,nozari2007thermodynamics}. Besides these ones, thermal aspects of field theories have also been calculated in different scenarios \cite{aa1,aa2,aa3,aa4,aa5,aa6,aa7,aa8,aa9,aa10,aa11,aa12,aa13,aa14}.

The non--commutativity of spacetime is a fundamental concept in modern theoretical physics, described by the relation $[x^\mu,x^\nu]=i \Theta^{\mu \nu}$, where $x^\mu$ is the spacetime coordinates and $\Theta^{\mu \nu}$ is an anti--symmetric constant tensor. Different methods have been proposed to incorporate non--commutativity into theories of gravity, including the use of the NC gauge de Sitter (dS) group, SO(4,1), in conjunction with the Poincaré group, ISO(3,1), through the Seiberg-–Witten (SW) map approach. This formalism has been employed to obtain a deformed metric for the Schwarzschild BH by Chaichian et al. \cite{chaichian2008corrections}. Alternatively, Nicolini et al. \cite{nicolini2006noncommutative} showed that the NC effect can also be modeled as an effect on the matter source term without altering the Einstein tensor part of the field equation. This can be achieved by replacing the point--like mass density on the right--hand side of the Einstein equation with a Gaussian smeared or Lorentzian distribution, with ${\rho _\Theta } = M{(4\pi \Theta )^{ - \frac{3}{2}}}e^{- \frac{{{r^2}}}{{4\Theta }}}$ and ${\rho _\Theta } = M\sqrt \Theta {\pi ^{ - \frac{3}{2}}}{({r^2} + \pi \Theta )^{ - 2}}$, respectively.

On the contrary, recent data from the Event Horizon Telescope (EHT) has ushered in a new era of exploration into the shadows and gravitational lensing around black holes, attracting significant attention. For example, in reference \cite{pulicce2023constraints}, the authors introduced Symmergent gravity, and its charged variant was subsequently derived and examined, with a focus on its properties, such as the Hawking temperature. Furthermore, this study delved into the constraints imposed on Symmergent gravity by analyzing its shadow properties in conjunction with data from EHT. Subsequently, we applied these findings to the weak field regime, utilizing the Gauss--Bonnet theorem to obtain the weak deflection angle.

Additionally, in the same metric context, quasinormal modes and greybody factors were investigated using the WKB approach, as detailed in \cite{gogoi2023quasinormal}. The influence of Symmergent gravity on null geodesics, which in turn affects the shadow radius and associated observables, was explored \cite{pantig2023testing}. Moreover, other important effects on the shadow and certain thermodynamic properties have been examined \cite{ccimdiker2021black}.

On top of that, the BH shadows were investigated in asymptotically safe gravity  \cite{lambiase2023investigating}. It was also investigated the \textit{quasinormal} modes of BHs in $f(Q)$ gravity \cite{gogoi2023quasinormal}. In \cite{pantig2023black}, the authors considered a spacetime metric for a BH surrounded by wave dark matter, and then they found the shadow radius of this setup. In general, some properties such as shadow, lensing, quasinormal modes, greybody bounds, and neutrino propagation by dyonic modified Maxwell BHs were studied in \cite{pantig2022shadow}.

Furthermore, it was scrutinized a model to examine the influence of torsion on the properties of charged black holes, including their shadow, deflection angle, and greybody radiation \cite{pantig2023testing}. This study was conducted in the context of both $M87^*$ and $Sgr. A^*$, utilizing data from the Event Horizon Telescope. Apart from that, the same characteristics were assessed for black holes governed by non--linear electrodynamics, along with a thin accretion disk \cite{uniyal2023probing}.



In \cite{yang2023probing}, the authors focused on the exploration of hairy black holes generated through gravitational decoupling. They analyzed the time--domain profiles of massless scalar fields, electromagnetic fields, and axial gravitational perturbations within these black hole spacetimes. The study also involved the determination of quasinormal mode (QNM) frequencies using the Prony method, 6th and 13th-order WKB approximations, and the discussion of greybody factors and high--energy absorption cross--sections, employing the sinc approximation.

In particular, it was first first examined a non-spinning black hole metric within the Dehnen profile \cite{pantig2022dehnen}. The authors then extended this study to include spin parameters, investigating their effects on various black hole properties, such as the horizon structure, ergoregions, geodesics, shadow, and weak deflection angle. 

In the context of modified theories of gravity, the BH shadows have been found for a variety of different models. In particular,  \cite{ovgun2020testing} introduced a new spherically symmetric solution in the generalized Einstein--Cartan--Kibble--Sciama gravity theory, which was used to investigate weak gravitational lensing and its resulting shadow. Furthermore, in \cite{ovgun2018shadow}, the authors computed the shadow and deflection angle for the Kerr--Newman--Kasuya spacetime.

This paper is organized as follows: in Sec. \ref{sec2}, we briefly
review the framework of implementing the non--commutativity through the Schwarzschild metric. After that, we propose a new approach via the mass deformation for doing so. In Sec. \ref{sec3}, we calculate the thermodynamic properties of the NC BH and compare our results with the previous results encountered in the literature. In Sec. \ref{sec4}, we provide the \textit{quasinormal} frequencies of such a BH hole using fundamentally the methods: WKB approximation and Pöschl--Teller fitting approach. In Sec. \ref{sec5}, we study the grey--body factor and the absorption cross section. In Sec. \ref{sec6}, we examine the geodesic path of massless and massive particles; also, we display the effect of the non--commutativity to the photon sphere. In Sec. \ref{sec7}, we demonstrate how the shadow radius depends on the $\Theta$ parameter of non--commutative theory and provide some constraints on $\Theta$ applying the observation of Sgr A* from EHT. In Sec. \ref{sec8}, we investigate the deflection angle.
Finally, \ref{conclusion}, we exhibit our conclusion.

\section{SCHWARZSCHILD BLACK HOLE WITH DEFORMED MASS} \label{sec2}

In Ref. \cite{9}, the authors constructed a deformation of the gravitational field by gauging the NC de Sitter SO(4,1) group and using the Seiberg--Witten map. They calculated the deformed gravitational gauge potentials (tetrad fields) $ \hat e_\mu ^a(x,\Theta )$ by contracting the NC group SO(4,1) to the Poincaré one ISO(3,1). They applied these potentials to the Schwarzschild BH and arrived at a deformed Schwarzschild metric with a non--commutativity parameter up to the second order, given as follows:
\begin{equation}\label{met}
\begin{array}{l}
{{\hat g}_{00}} = {g_{00}} - \frac{{\alpha (8r - 11\alpha )}}{{16{r^4}}}{\Theta ^2} + \mathcal{O}({\Theta ^4}),\\
{{\hat g}_{11}} = {g_{11}} - \frac{{\alpha (4r - 3\alpha )}}{{16{r^2}{{(r - \alpha )}^2}}}{\Theta ^2} + \mathcal{O}({\Theta ^4}),\\
{{\hat g}_{22}} = {g_{22}} - \frac{{2{r^2} - 17\alpha r + 17{\alpha ^2}}}{{32r(r - \alpha )}}{\Theta ^2} + \mathcal{O}({\Theta ^4}),\\
{{\hat g}_{33}} = {g_{33}} - \frac{{({r^2} + \alpha r - {\alpha ^2})\cos \theta  - \alpha (2r - \alpha )}}{{16r(r - \alpha )}}{\Theta ^2} + \mathcal{O}({\Theta ^4}).
\end{array}
\end{equation}
In the context of this study, the parameter $ \alpha $ is defined as $ \alpha  = \frac{{2GM}}{{{c^2}}} $, where $G$ is the gravitational constant, $M$ is the mass of the BH, and $c$ is the speed of light. The symbol $\hat{g}_{\mu \nu}$ represents the deformed metric tensor in spherical coordinates, while $g_{\mu \nu}$ stands for the well--known Schwarzschild metric. To obtain the deformed Schwarzschild event horizon radius, we set $\hat{g}_{00}$ equal to zero, as done in Ref. \cite{araujo2023thermodynamics}. This leads to the following expression:
\begin{equation}\label{rad}
{r_{s\Theta }} = 2M + \frac{{3{\Theta ^2}}}{{32M}},
\end{equation}
where $r_s=2M$ is the known Schwarzschild BH. We attribute the deformed NC mass of the Schwarzschild BH to its radius as $r_{s\Theta}=2M_\Theta$, which entails to a newly defined deformed mass as:
\begin{equation}\label{mass}
\ {M_\Theta } = M + \frac{{3}}{{64M}}{\Theta ^2}.
\end{equation}
The standard Schwarzschild metric is used in this work with a deformed NC mass Eq. (\ref{mass}). This approach allows us to investigate the thermodynamic properties of the Schwarzschild BH and we compare our results with previous studies concerning the metric deformation instead \cite{araujo2023thermodynamics}. As it is straightforward to note, when the NC parameter approaches zero, the well--known Schwarzschild mass is obtained. 


\section{Thermodynamic properties of non--commutative Schwarzschild black hole with deformed mass\label{sec3}}

We begin our analysis by focusing on the temperature of the NC Schwarzschild BH. To do so, we consider the deformed event horizon radius, $r_{h\Theta}$, in the well--known Hawking temperature relation for Schwarzschild BH, given by $T=\frac{1}{8\pi M}$. Using Eq. (\ref{mass}), the modified temperature reads:
\begin{equation}\label{R0}
\ {T_\Theta } = \frac{1}{{4\pi \sqrt {{g_{00}}{g_{11}}} }}{\left. {\frac{{\mathrm{d}{g_{00}}}}{{\mathrm{d}r}}} \right|_{r = {r_{h\Theta }}}} \approx \frac{1}{{8\pi M}}\left(1 - \frac{{3}}{{64{M^2}}}{\Theta ^2}\right).
\end{equation}

\begin{figure}
	\centering
	\includegraphics[width=80mm]{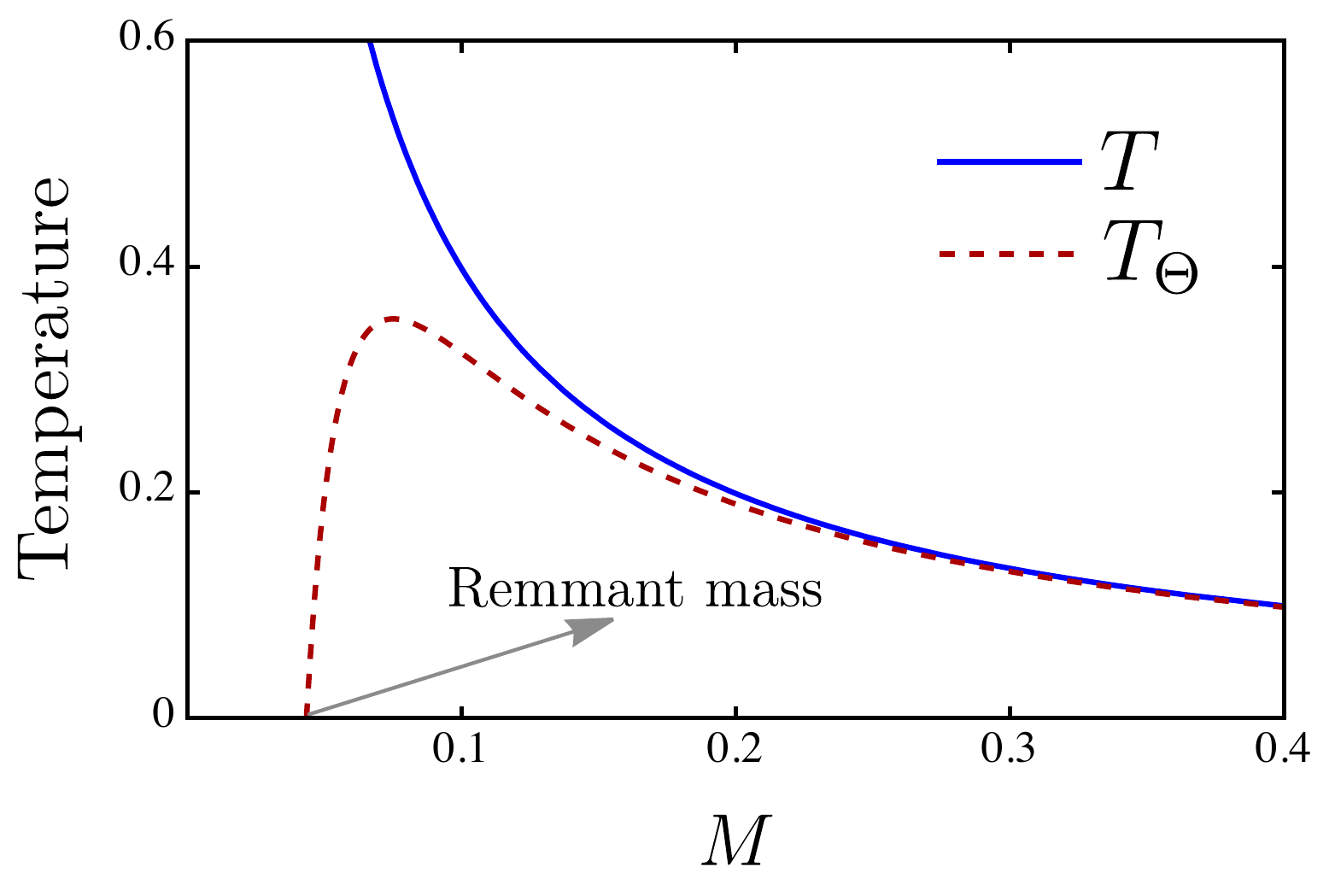}
	\caption{The original temperature and modified NC temperature as functions of mass for a Schwarzschild BH. The NC curve corresponds to a NC spacetime with $\Theta^2=0.04$.}
	\label{fig:Tem}
\end{figure}

It implies that the NC temperature is dependent on the NC parameter as $\lim\limits_{\Theta^2\to 0} T_\Theta =T$. To examine the effect of $\Theta$ on this thermal quantity, we have plotted the temperature versus mass for $\Theta^2=0$ (the usual case of Schwarzschild BH) and $\Theta^2=0.04$ in Fig. \ref{fig:Tem}. The figure demonstrates that non--commutativity induces a substantial change in temperature. In contrast to the Schwarzschild BH, in our case, there exists a minimal non--zero mass at the final stage of BH evaporation. The value of this remnant mass depends on the $\Theta$ value.

As the metric parameter does not change in this approach, we can find the horizon area by using the standard Schwarzschild--like metric. The area of the event horizon can be written as:

\begin{equation}
\ {A_\Theta } =\int {\int {\sqrt {{g_{22}}{g_{33}}} } } \mathrm{d}\theta \mathrm{d}\varphi = 4\pi r_{h\Theta }^2.
\end{equation}
By using it, we can calculate the deformed entropy of the BH as follows:
\begin{equation}\label{ent}
\ {S_\Theta } = \frac{{{A_\Theta }}}{4} \approx 4\pi {M^2} - \frac{{3\pi}}{{8}}{\Theta ^2}.
\end{equation}
Furthermore, the heat capacity can also be addressed to:
\begin{equation}
\ {C_{V\Theta }} = {T_\Theta }\frac{{\partial {S_\Theta }}}{{\partial {T_\Theta }}} = {T_\Theta }\frac{{{{\partial {S_\Theta }} \mathord{\left/
				{\vphantom {{\partial {S_\Theta }} {\partial M}}} \right.
				\kern-\nulldelimiterspace} {\partial M}}}}{{{{\partial {T_\Theta }} \mathord{\left/
				{\vphantom {{\partial {T_\Theta }} {\partial M}}} \right.
				\kern-\nulldelimiterspace} {\partial M}}}} =  - 8\pi {M^2}\left(\frac{{1 - \frac{3}{{64{M^2}}}{\Theta ^2}}}{{1 - \frac{9}{{64{M^2}}}{\Theta ^2}}}\right).	
\end{equation}

\begin{table}[t]
	\centering
		\caption{\label{thermo}Comparison of the thermodynamic properties between a NC Schwarzschild BH obtained by the present study (via deformed mass) and the previous results (from deformed metric).}
	\setlength{\arrayrulewidth}{0.3mm}
	\setlength{\tabcolsep}{30pt}
	\renewcommand{\arraystretch}{1}
	\begin{tabular}{c c c}
		\hline \hline
		~ & Deformed mass & Deformed metric \\ \hline
		$T_\Theta$ & $\frac{1}{{8\pi M}}\left(1 - \frac{{3}}{{64{M^2}}}{\Theta ^2}\right)$ & $\frac{1}{{8\pi M}} - \frac{{3}}{{512{M^2}}}{\Theta ^2}$ \\
		$A_\Theta$ & $4\pi r_{s}^{2}$+$\frac{3\pi }{2}\Theta^2$ &$4\pi r_{s}^{2}$+$\frac{5\pi }{16}\Theta^2$  \\ \
		$S_\Theta$ &  $4\pi {M}^2$+$\frac{3\pi }{8}\Theta^2$  & $4\pi {M}^2$+$\frac{5\pi }{64}\Theta^2$ \\ 
		$C_{V\Theta}$ &  $- 8\pi {M^2}\left(\frac{{1 - \frac{3}{{64{M^2}}}{\Theta ^2}}}{{1 - \frac{9}{{64{M^2}}}{\Theta ^2}}}\right)$ & $ - 8\pi {M^2}\left(\frac{{64{M^2}-3{\Theta ^2}}}{{64{M^2} -9{\Theta ^2}}}\right)$ \\ \hline\hline
	\end{tabular}
\end{table}

Table \ref{thermo} displays the thermodynamic properties of the NC Schwarzschild BH obtained by using two different methods: mass and metric deformation. The first column corresponds to the results obtained by considering a known Schwarzschild metric with a NC deformed mass given by Eq. (\ref{mass}), while the second one accounts for the properties found using a deformed Schwarzschild metric.


\section{QUASINORMAL FREQUENCIES IN NON--COMMUTATIVE MODEL 
	\label{sec4}}

During the ringdown phase of a BH merger for instance, we can infer about the \textit{quasinormal} modes. These ones are unique oscillation patterns of the system and are independent of the initial perturbations. The reason for this is that they represent the free oscillations of the spacetime, which do not rely on the initial conditions. In contrast to the \textit{normal} modes, the \textit{quasinormal} modes correspond to an open system, resulting in the dissipation of energy through the emission of gravitational waves. These modes can be described as poles of the complex Green function. To calculate the \textit{quasinormal} frequencies, we need to find wave equation solutions for a system governed by a background metric $g_{\mu\nu}$. However, obtaining analytical solutions for such modes is typically not feasible.

Various methods have been proposed in the literature to obtain \textit{analytical} solutions for the \textit{quasinormal} modes. One of the most well--known approaches is the WKB (Wentzel--Kramers--Brillouin) method, which was first introduced by Will and Iyer \cite{iyer1987black,iyer1987black1}. Later, Konoplya improved this method to the sixth order \cite{konoplya2003quasinormal}. In our calculations, we consider perturbations through the scalar field. Therefore, we can express the Klein--Gordon equation in a curved spacetime as follows:
\begin{equation}
\frac{1}{\sqrt{-g}}\partial_{\mu}(g^{\mu\nu}\sqrt{-g}\partial_{\nu}\Phi) = 0 \label{KL},
\end{equation}
where we shall use the NC Schwarzschild BH metric constructed by applying the modified NC mass 
\begin{equation}\label{metric}
 d{s^2} =  - {f_\Theta }(r)d{t^2} + f_\Theta ^{ - 1}(r)d{t^2} + {r^2}d{\Omega ^2},
\end{equation}
with the $f_\Theta$ being determined by
\begin{equation}
 {f_\Theta }(r) = 1 - \frac{{2{M_\Theta }}}{r}.
\end{equation}

\begin{figure}
	\centering
	\includegraphics[width=90mm]{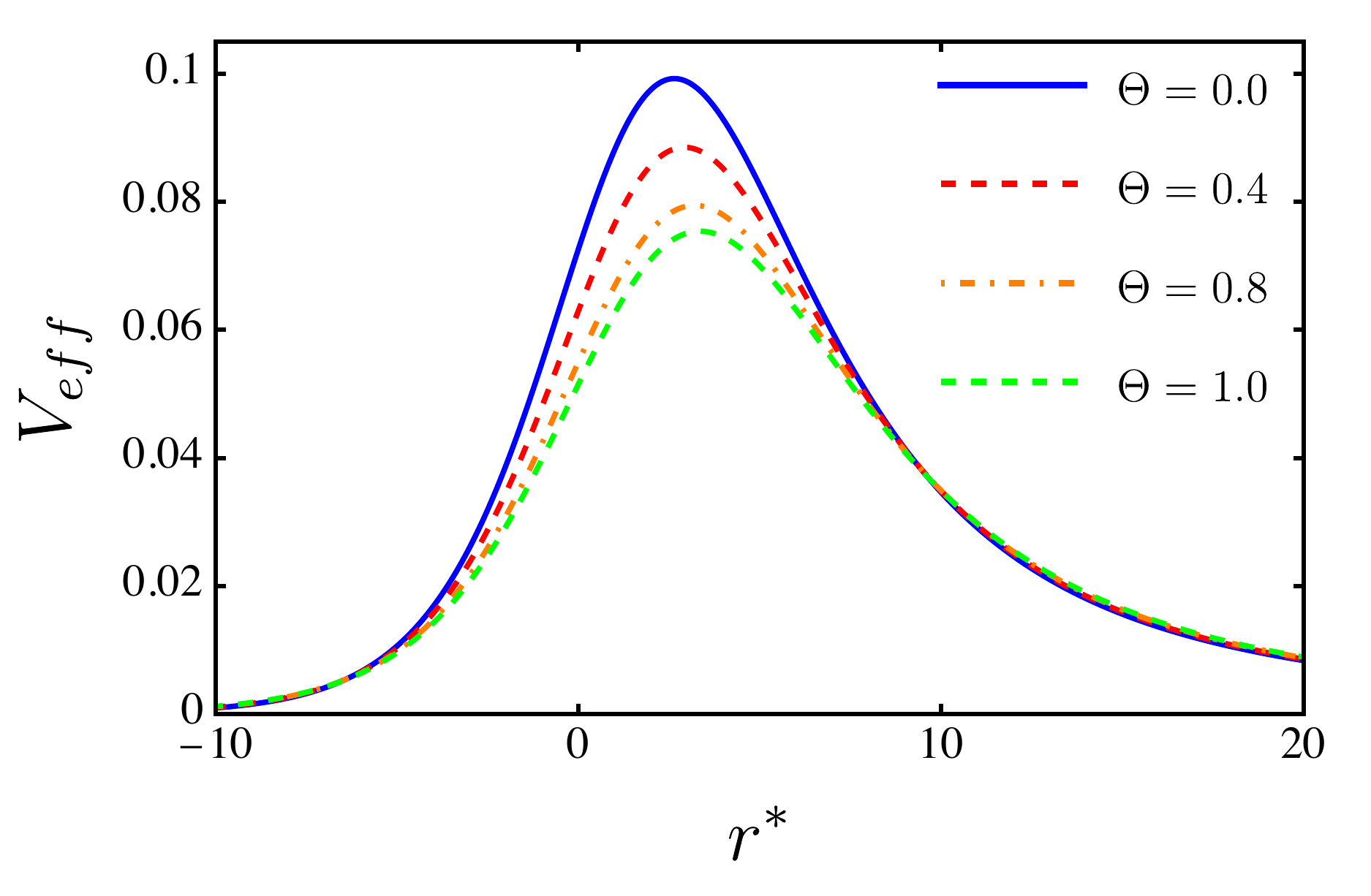}
	\caption{Effective potential for scalar field with $M = 0.5$, $l=1$ and, concerning different values of $\Theta^2$ }
	\label{fig:Veff}
\end{figure}

In this work, we focus on studying the scalar field perturbations. In addition, we take advantage of spherical symmetry by decomposing it in the following manner:
\begin{equation}
\Phi(t,r,\theta,\varphi) = \sum_{l=0}^{\infty}\sum_{m=-l}^{l}r^{-1}\Psi_{lm}(t,r)Y_{lm}(\theta,\varphi), \label{decomposition}
\end{equation}
where $Y_{lm}(\theta,\varphi)$ denotes the spherical harmonics. Substituting the decomposition encountered in Eq. (\ref{decomposition}) into Eq. (\ref{KL}), we shall have a wave--like equation:
\begin{equation}
-\frac{\partial^{2} \Psi}{\partial t^{2}}+\frac{\partial^{2} \Psi}{\partial r^{*2}} + V_{eff}(r)\Psi = 0,\label{schordingereq}
\end{equation}
where we introduce the tortoise coordinate $r^{*}$, which runs from $-\infty$ to $+\infty$ all over the spacetime. We define it as $\mathrm{d} r^{*} = \sqrt{[1/f_{\Theta}(r)]}\mathrm{d}r$. Although the \textit{backreaction} effects are interesting to study, we do not address this feature in this manuscript. Here, $V_{eff}(r)$ denotes the \textit{Regge--Wheeler} potential or effective potential, encoding the geometry of the BH given by 
\begin{equation}
V_{eff}(r) =  \frac{{f_\Theta(r)}}{r}\frac{{\mathrm{d}f_\Theta}}{{\mathrm{d}r}} + f_\Theta(r)\frac{{l(l + 1)}}{{{r^2}}}=\left( 1 - \frac{2M_\Theta}{r}\right)
\left( \frac{{l(l + 1)}}{{{r^2}}}+\frac{2M_\Theta}{r^3}\right). \label{effectiveepotential}
\end{equation}
To visualize such a potential in tortoise coordinate for various NC parameters, we display Fig. \ref{fig:Veff}.
It illustrates a notable trend caused by the non--commutativity of the spacetime: as $\Theta^2$ increases, the maximum value of the effective potential decreases. In the next section, we will effectively perform calculations to determine the \textit{quasinormal} modes of the scalar field in the presence of non--commutativity.

Notice that the Schrödinger--like equation (\ref{schordingereq}) cannot be solved \textit{analytically}, so that extensive \textit{analytical} and \textit{numerical} methods have been proposed to calculate the \textit{quasinormal} modes from BHs. One of the most effective methods is to approximate the effective potential analytically by an exactly solvable potential. Several semi--analytical approaches have been suggested to find the QNMs, including the methods proposed by Mashhoon and others \cite{19,23,25}. Mashhoon's technique is one of the simplest ones, where the effective potential is substituted with the Pöschl--Teller potential \cite{19}. within another approach, the such a potential is replaced by the Rosen--Morse function, which gives an analytical solution for the radial wave function \cite{22}. The continued fraction method is another well--known procedure that was proposed by Leaver \cite{24,33,34}.

Our analysis employs both the Pöschl--Teller fitting approach and the 6th order WKB method to examine the QNMs of a Schwarzschild--like BH in the context of a NC theory via the mass deformation. Through this approach, we strive to gain a deeper understanding of the impact of noncommutativity on the system.


\subsection{Quasinormal modes via WKB method}

The objective of this section is to derive stationary solutions, which are obtained under the assumption that $\Psi(t,r)$ follows the form $e^{-i\omega t} \psi(r)$, where $\omega$ represents frequency. Using this assumption, the time-independent part of Eq. (\ref{schordingereq}) can be separated into the following form:
\begin{equation}
\frac{\partial^{2} \psi}{\partial r^{*2}} - \left[ \omega^{2} - V_{eff}(r)\right]\psi = 0.
\label{timeindependent}
\end{equation}
To solve this equation, appropriate boundary conditions must be considered. In this case, acceptable solutions are those that are purely ingoing near the horizon:
\[
    \psi^{\text{in}}(r^{*}) \sim 
\begin{cases}
    C_{l}(\omega) e^{-i\omega r^{*}} & ( r^{*}\rightarrow - \infty)\\
    A^{(-)}_{l}(\omega) e^{-i\omega r^{*}} + A^{(+)}_{l}(\omega) e^{+i\omega r^{*}} & (r^{*}\rightarrow + \infty),\label{boundaryconditions11}
\end{cases}
\]
Here, $C_{l}(\omega)$, $A^{(-)}_{l}(\omega)$, and $A^{(+)}_{l}(\omega)$ are complex constants. The \textit{quasinormal} modes of a BH are defined by the set of frequencies ${\omega_{nl}}$, such that $A^{(-)}_{l}(\omega_{nl})=0$. This condition ensures that the modes are related to a purely outgoing wave at spatial infinity and a purely ingoing wave at the event horizon. The integers $n$ and $l$ are known as the overtone and multipole numbers, respectively. The spectrum of \textit{quasinormal} modes is determined by the eigenvalues of Eq. (\ref{timeindependent}). To analyze these modes, we employ the WKB method, which is a semi-analytical approach that takes advantage of the analogy with quantum mechanics.

The WKB approximation was first used by Schutz and Will to compute \textit{quasinormal} modes in the study of scattering particles around BHs \cite{schutz1985black}. Subsequently, Konoplya further developed this technique \cite{konoplya2003quasinormal,konoplya2004quasinormal}. However, for this method to be valid, the potential must have a barrier-like shape that approaches constant values as $r^{*} \rightarrow \pm \infty$. The \textit{quasinormal} modes can then be obtained by fitting the power series of the solution near the maximum of the potential at its turning points \cite{santos2016quasinormal}.

The formula for the \textit{quasinormal} modes, as developed by Konoplya, is expressed as:
\begin{equation}
\frac{i(\omega^{2}_{n}-V_{0})}{\sqrt{-2 V^{''}_{0}}} - \sum^{6}_{j=2} \Lambda_{j} = n + \frac{1}{2},
\end{equation}
where $V^{''}_{0}$ denotes the second derivative of the potential at its maximum $r_{0}$, and $\Lambda_{j}$ represents constants that rely on the effective potential and its derivatives at the maximum. It is worth noting that the WKB approximation has been extended up to the $13$th order, as recently proposed by Matyjasek and Opala \cite{matyjasek2017quasinormal}.



\subsection{Quasinormal modes via P\"{o}sch--Teller fitting method}

In the Pöschl--Teller fitting method, the effective potential is approximated with the Pöschl--Teller function \cite{23}, which provides an analytical solution for Eq. (\ref{effectiveepotential}). The effective potential is substituted with the Pöschl--Teller function in the form:
\begin{equation}
V_{eff} \sim {V_{PT}} = \frac{{{V_0}}}{{{{\cosh }^2}\alpha ({r^{*}} - r_0^{*})}}
\end{equation}
where $V_0$ and $\alpha$ denote the height and curvature of the potential at its maximum, respectively. However, these values are obtained through a fitting method, as outlined in Ref. \cite{39}, and we follow the same approach. By substituting the effective potential with the Pöschl--Teller function, the QNMs are evaluated using the following formula:
\begin{equation}\label{Mash}
\omega_{n} = i\alpha \left(n + \frac{1}{2}\right) \pm \alpha \sqrt {\frac{{{V_0}}}{{{\alpha ^2}}} - \frac{1}{4}}.
\end{equation}

We applied this equation to calculate the QNMs for different multipoles and NC parameter values, in order to investigate the effects of NC space. Overall, the Pöschl-Teller fitting method approximates the effective potential with a function that can be analytically solved, allowing us to calculate the QNMs using a simple formula.

The calculated QNMs with different NC parameter values are presented in Tables \ref{table:pt} and \ref{table:wkb} for the Pöschl-Teller fitting method and WKB method, respectively. QNMs are composed of two main components: the real part, which represents the actual frequency of the oscillation, and the imaginary part, which is linked to the damping timescale and can be used to investigate the stability of the BH.

In our analysis, we considered three families of multipoles with $l = 0, 1, 2$, where monopoles satisfy ($n<l$) and $\Theta$ is varied. The case of $\Theta=0$ corresponds to the original Schwarzschild BH. In both methods, we observed that increasing the NC parameter results in a decrease in the propagating frequency ($\omega_{ln}$) when comparing the real part of the frequencies. However, the imaginary part of the frequency decreases as the $\Theta$ value increases, indicating that higher values of the NC parameter lead to greater stability for the BH.  
\begin{table}[!ht]
	\centering
		\caption{\label{table:pt}Comparing the QNMs of scalar perturbations in NC space for a Schwarzschild BH with a mass of $M=0.5$, obtained for different $\Theta^2$ values using the PT fit method.}
	\setlength{\arrayrulewidth}{0.5mm}
	\renewcommand{\arraystretch}{1}
	\begin{tabular}{|l|l|l|l|l|l|}
		\hline\hline
		\multicolumn{2}{|c|}{$l$~~~~$n$} & $\Theta^2=0$ &  $\Theta^2=0.4$ &  $\Theta^2=0.8$ &  $\Theta^2=1$ \\ \hline\hline
		\multicolumn{2}{|c|}{$0 ~~~~~ 0$} & 0.231827-0.22726i & 0.215971-0.211033i & 0.201549-0.197664i & 0.195518-0.191031i \\ \hline
		\multicolumn{2}{|c|}{$1 ~~~~~ 0$} & 0.598404-0.194272i & 0.556588-0.181234i & 0.520331-0.168926i & 0.50381-0.164391i \\ \hline
		\multicolumn{2}{|c|}{$~ ~~~~~ 1$}& 0.598404-0.582815i & 0.556588-0.543701i & 0.520331-0.506778i & 0.50381-0.493172i \\ \hline
  		\multicolumn{2}{|c|}{$2 ~~~~~ 0$} & 0.974511-0.188406i & 0.906435-0.174858i & 0.84743-0.164313i & 0.820557-0.158078i \\ \hline
		\multicolumn{2}{|c|}{$~ ~~~~~ 1$}  & 0.974511-0.565219i & 0.906435-0.524574i & 0.84743-0.492939i & 0.820557-0.474233i \\ \hline
		\multicolumn{2}{|c|}{$~ ~~~~~2$} & 0.974511-0.942031i & 0.906435-0.87429i & 0.84743-0.821565i & 0.820557-0.790389i \\ \hline
	\end{tabular}
\end{table}
\begin{table}[!ht]
	\centering
	\caption{\label{table:wkb}Comparison of the QNMs of scalar perturbations in NC Schwarzschild BHs for $M=0.5$ and different $\Theta^2$ values using the WKB method.}
	\setlength{\arrayrulewidth}{0.5mm}
	\renewcommand{\arraystretch}{1}
	\begin{tabular}{|l|l|l|l|l|l|}
		\hline\hline
		\multicolumn{2}{|c|}{$l$~~~~$n$} & $\Theta^2=0$ &  $\Theta^2=0.4$ &  $\Theta^2=0.8$ &  $\Theta^2=1$  \\ \hline\hline
			\multicolumn{2}{|c|}{$0 ~~~~~ 0$} & 0.22092-0.201645i & 0.205527-0.187559i, & 0.192133-0.175318i & 0.186048-0.169797i \\ \hline
			\multicolumn{2}{|c|}{$1 ~~~~~ 0$} & 0.585819-0.195523i & 0.544948-0.181882i & 0.509408-0.17002i & 0.493322-0.164651i \\ \hline
			\multicolumn{2}{|c|}{$~ ~~~~~ 1$} & 0.528942-0.613037i & 0.492039-0.570266i & 0.45995-0.533075i & 0.445425-0.516241i \\ \hline
			\multicolumn{2}{|c|}{$2 ~~~~~ 0$} & 0.967284-0.193532i & 0.899799-0.18003i & 0.841116-0.168289i & 0.814555-0.162974i \\ \hline
			\multicolumn{2}{|c|}{$~ ~~~~~ 1$} & 0.927694-0.591254i & 0.862971-0.550004i & 0.80669-0.514134i & 0.781216-0.497898i \\ \hline
		\multicolumn{2}{|c|}{$~ ~~~~~ 2$} & 0.860772-1.0174i & 0.800718-0.946419i & 0.748497-0.884696i & 0.72486-0.856758i \\ \hline
	\end{tabular}
\end{table}

\section{GREYBODY FACTOR AND ABSORPTION CROSS SECTION IN NON--COMMUTATIVE MODEL 
}\label{sec5}

In this section, we will briefly examine the scattering process using the WKB method. Another significant aspect of field perturbations around a BH spacetime is the absorption cross-section. The greybody factor, defined as the probability for an outgoing wave to reach infinity or the probability for an incoming wave to be absorbed by the BH \cite{30,31,32}, is crucial in determining the tunneling probability of the wave through the effective potential of the given BH spacetime. As shown in Fig. \ref{fig:Veff}, the effective potential is influenced by the $\Theta$ parameter. Thus, we aim to investigate the impact of non-commutativity on the greybody factor and absorption cross--section.

Scattering via the WKB method is an important aspect to consider and requires appropriate boundary conditions. In this case, we aim to determine the reflection and transmission coefficients, which are similar to those encountered in quantum mechanics for tunneling phenomena. To calculate these coefficients, we utilize the fact that $(\omega^{2}_{n}-V_{0})$ is purely real and obtain the following expression:

\begin{equation}
\Upsilon= \frac{i(\Tilde{\omega}^{2}-V_{0})}{\sqrt{-2 V''_{0}}} - \sum_{j=2}^{6} \Lambda_{j}(\Upsilon).
\end{equation}

The reflection and transmission coefficients can be obtained by analyzing the scattering through the semiclassical WKB approach, which has been recently studied in the literature \cite{konoplya2020quantum,konoplya2019higher,campos2022quasinormal}. The coefficients associated with the effective potential, denoted by $\Lambda_{j}(\Upsilon)$, are complex functions of the purely imaginary quantity $\Upsilon$, while the purely real frequency $\Tilde{\omega}$ is related to the quasinormal modes. The expressions for the reflection and transmission coefficients are given by:

\begin{equation}
|R|^{2} = \frac{|A_{l}^{(+)}|^{2}}{|A_{l}^{(-)}|^{2}} = \frac{1}{1+e^{-2i\pi \Upsilon}},
\end{equation}

\begin{equation}
|T|^{2} = \frac{|C_{l}|^{2}}{|A_{l}^{(-)}|^{2}} = \frac{1}{1+e^{+2i\pi \Upsilon}}.
\end{equation}

\begin{figure}
	\centering
		\includegraphics[width=80mm]{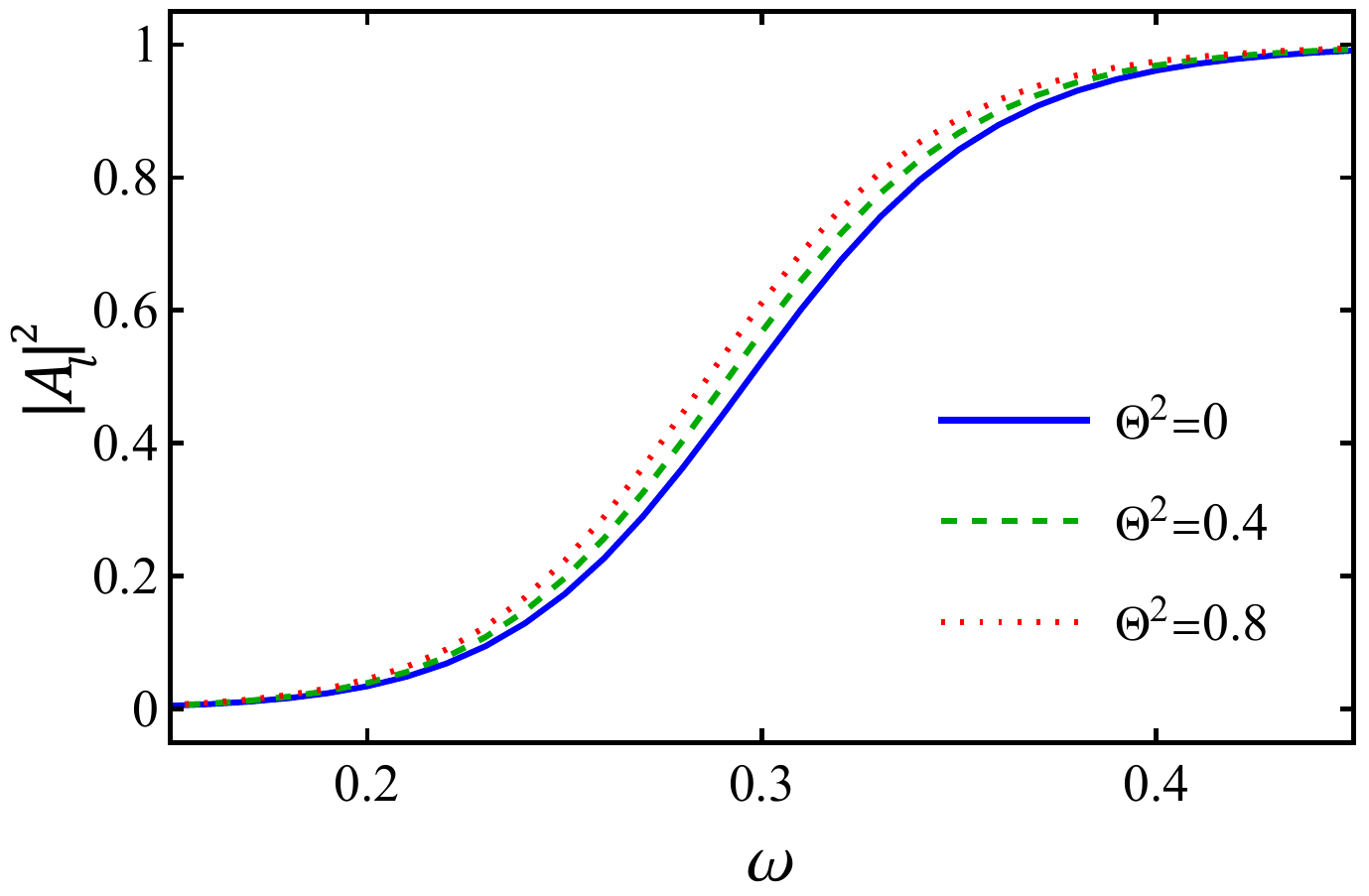}
		\hfill
		\includegraphics[width=80mm]{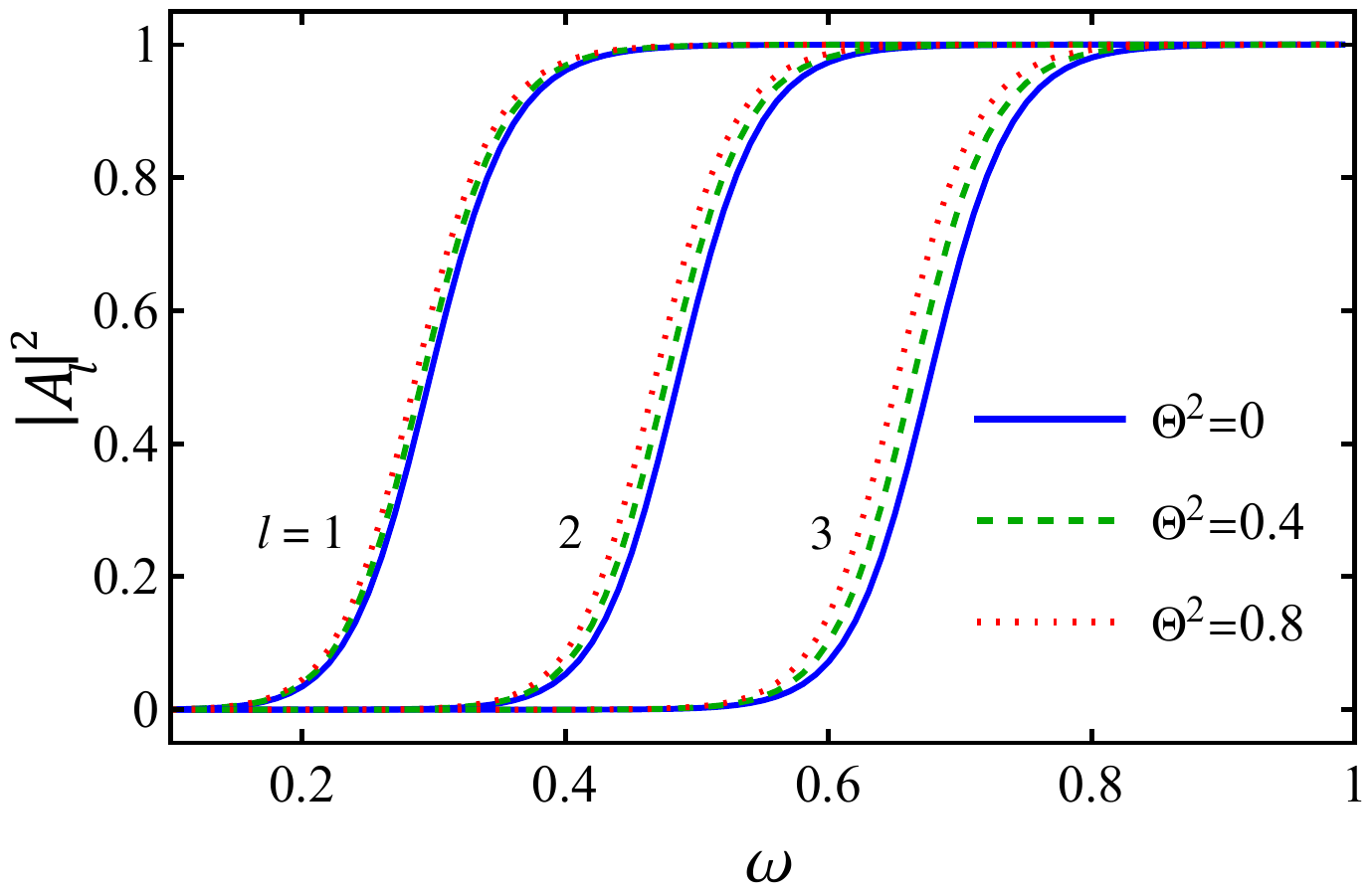}\\
		(a)~~~~~~~~~~~~~~~~~~~~~~~~~~~~~~~~~~~~~~~~~~~~~~~~~~~~(b)
		\caption{The grey-body factors of the scalar field are computed using the sixth-order WKB method. The left panel shows the results for $M=1$ and $l=1$, while the right panel shows the results for $M=1$ and $l=1$, $l=2$, and $l=3$ for various values of $\Theta$.}
		\label{fig:greybody}
\end{figure}

The Grey--body factors of the scalar field computed using the sixth order WKB method are shown in Fig. \ref{fig:greybody} for various values of the NC parameter, specifically for multipole $l=1$ and other values of $l$ with varying $\Theta$. As the figure illustrates, increasing the value of $\Theta$ leads to an increase in the Grey--body factors, indicating a greater fraction of the scalar field is penetrating the potential barrier. As the value of $\Theta$ increases, the potential barrier for scalar fields becomes lower, enabling more particles to transmit through the barrier. In contrast, the effective potential of the scalar field increases as $\Theta$ decreases, leading to a reduction in the Grey--body factor and the detection of a lower fraction of incoming flow of Hawking radiation by the observer.

The partial absorption cross section can be calculated by utilizing the transmission coefficient, which is defined as:
\begin{equation}
{\sigma _l} = \frac{{\pi (2l + 1)}}{{{\tilde{\omega}^2}}}{\left| {{T_l}(\tilde{\omega} )} \right|^2},
\end{equation}
where $l$ is the mode number and $\tilde{\omega}$ is the frequency. The total absorption cross section can be obtained by summing over all partial absorption cross sections, which can be expressed as:
\begin{equation}
{\sigma _{abs}} = \sum\limits_l {{\sigma _l}}.
\end{equation}

\begin{figure}[!]
	\centering
	\includegraphics[width=80mm]{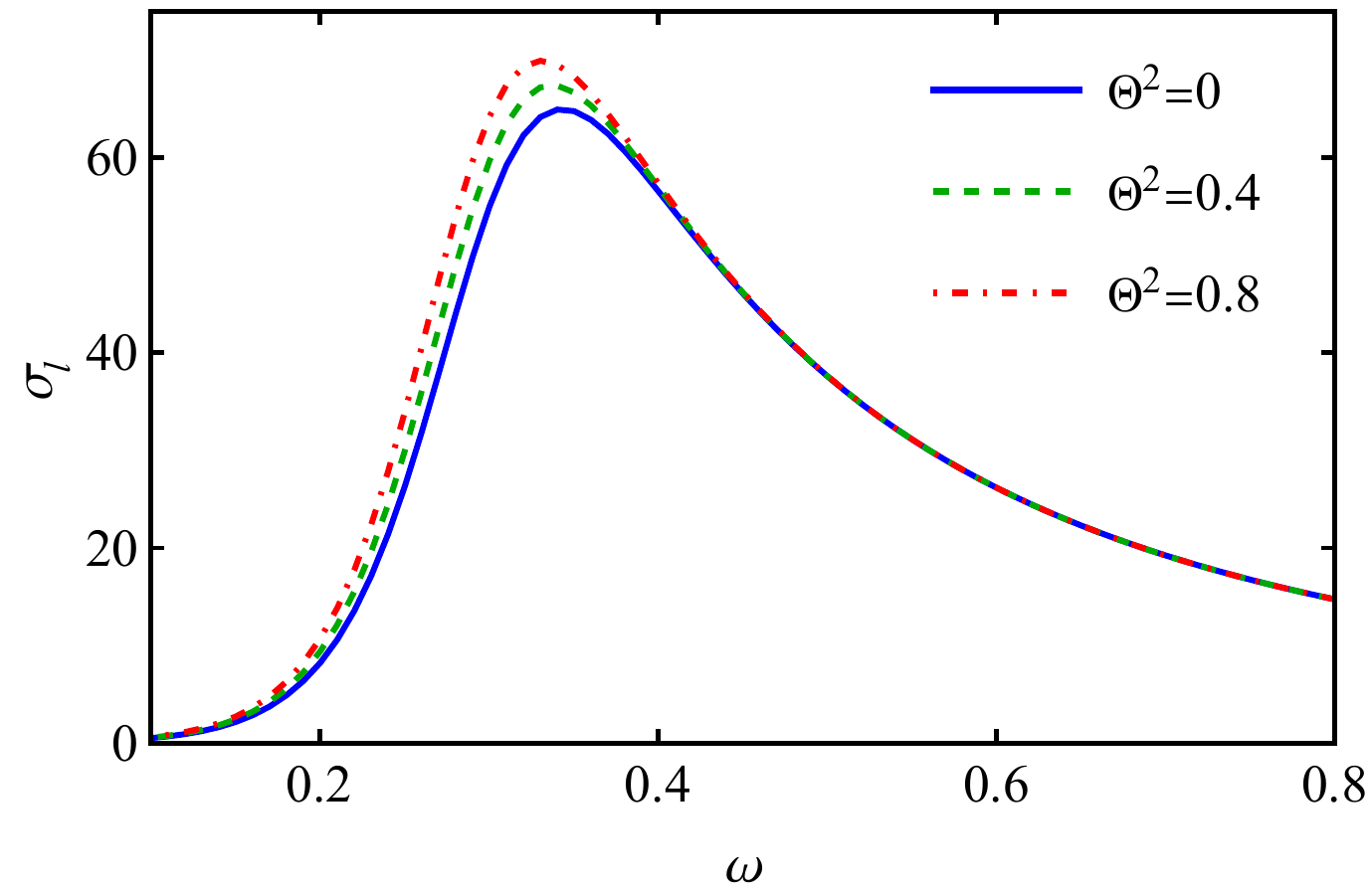}
	\hfill
	\includegraphics[width=80mm]{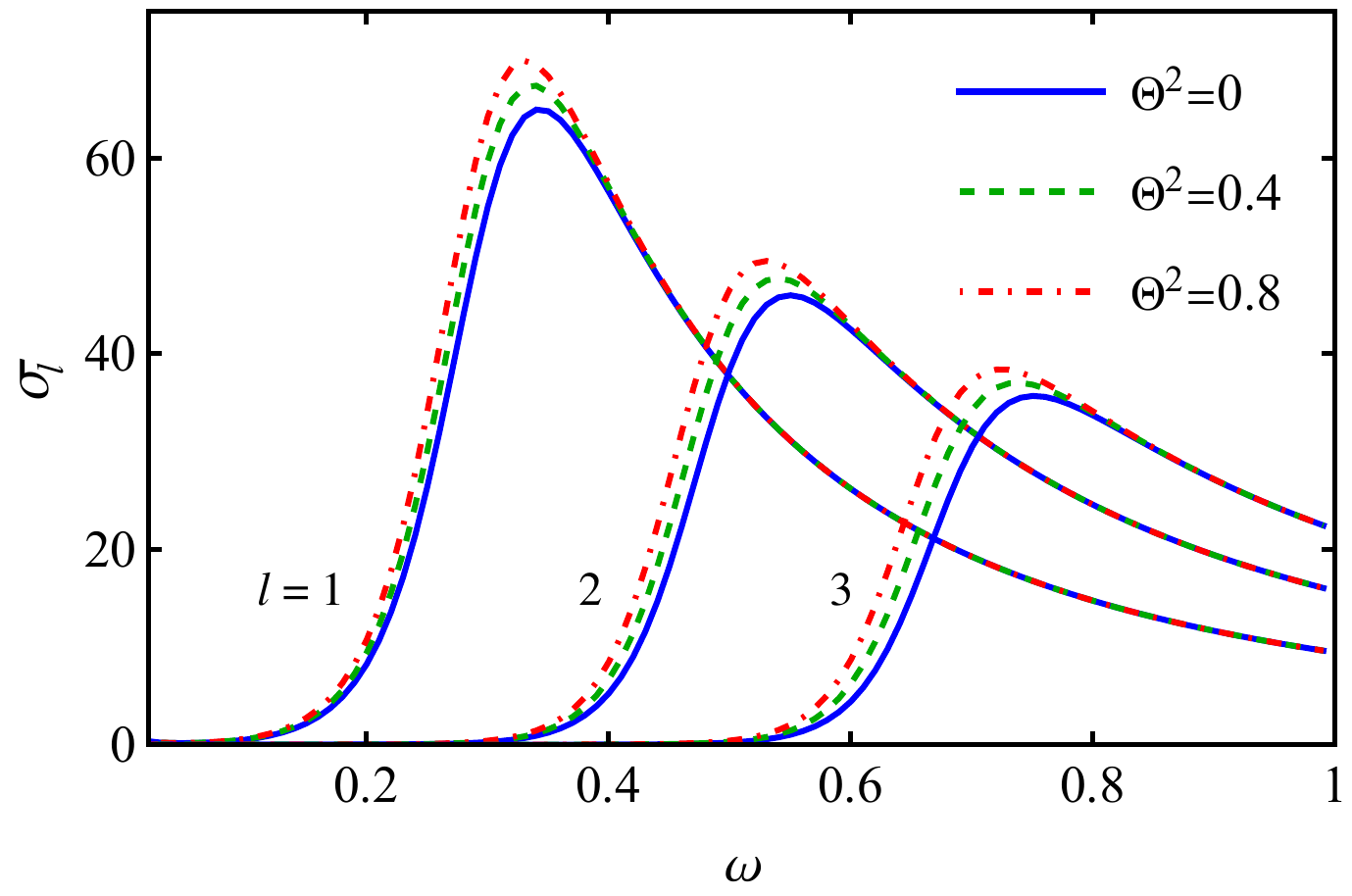}
	\caption{In the left plot, we can see the partial absorption cross section for the scalar field with mode $l=1$ of $M=1$ compared to NC BHs with $\Theta^2=0.4$ and $\Theta^2=0.8$. The right plot shows the partial absorption cross section for modes $l=1,2,3$ of the scalar field with $M=1$ and different values of $\Theta$.}
	\label{fig:cross1}
\end{figure}

\begin{figure}[!]
	\centering
	\includegraphics[width=80mm]{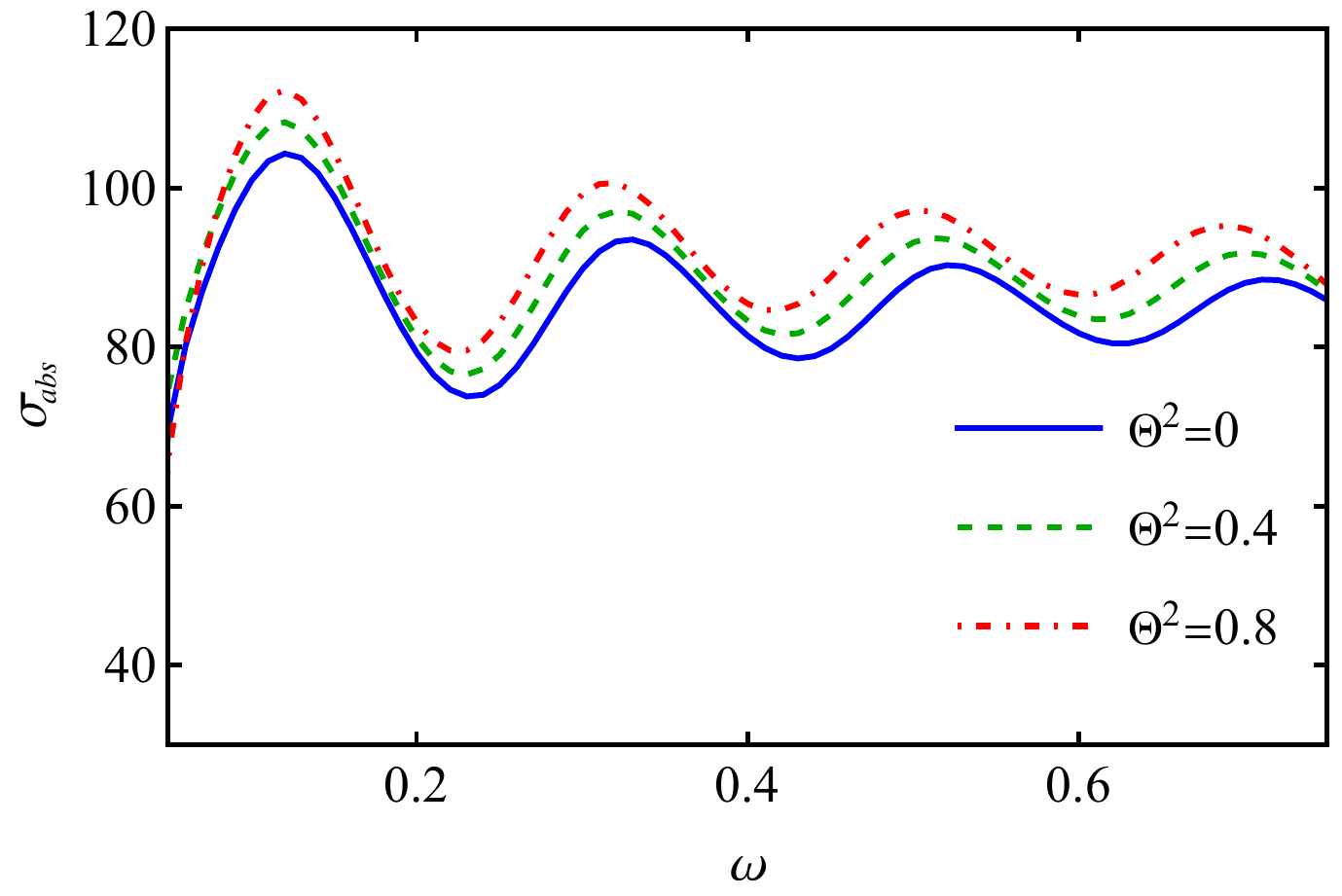}
	\caption{The total absorption cross section of massless scalar waves for a Schwarzschild BH with $M=1$ for monopoles ranging from $l=0$ to $l=3$ and different values of the non--commutativity parameter $\Theta$.}
	\label{fig:cross2}
\end{figure}

\begin{figure}
	\centering
	\includegraphics[width=80mm]{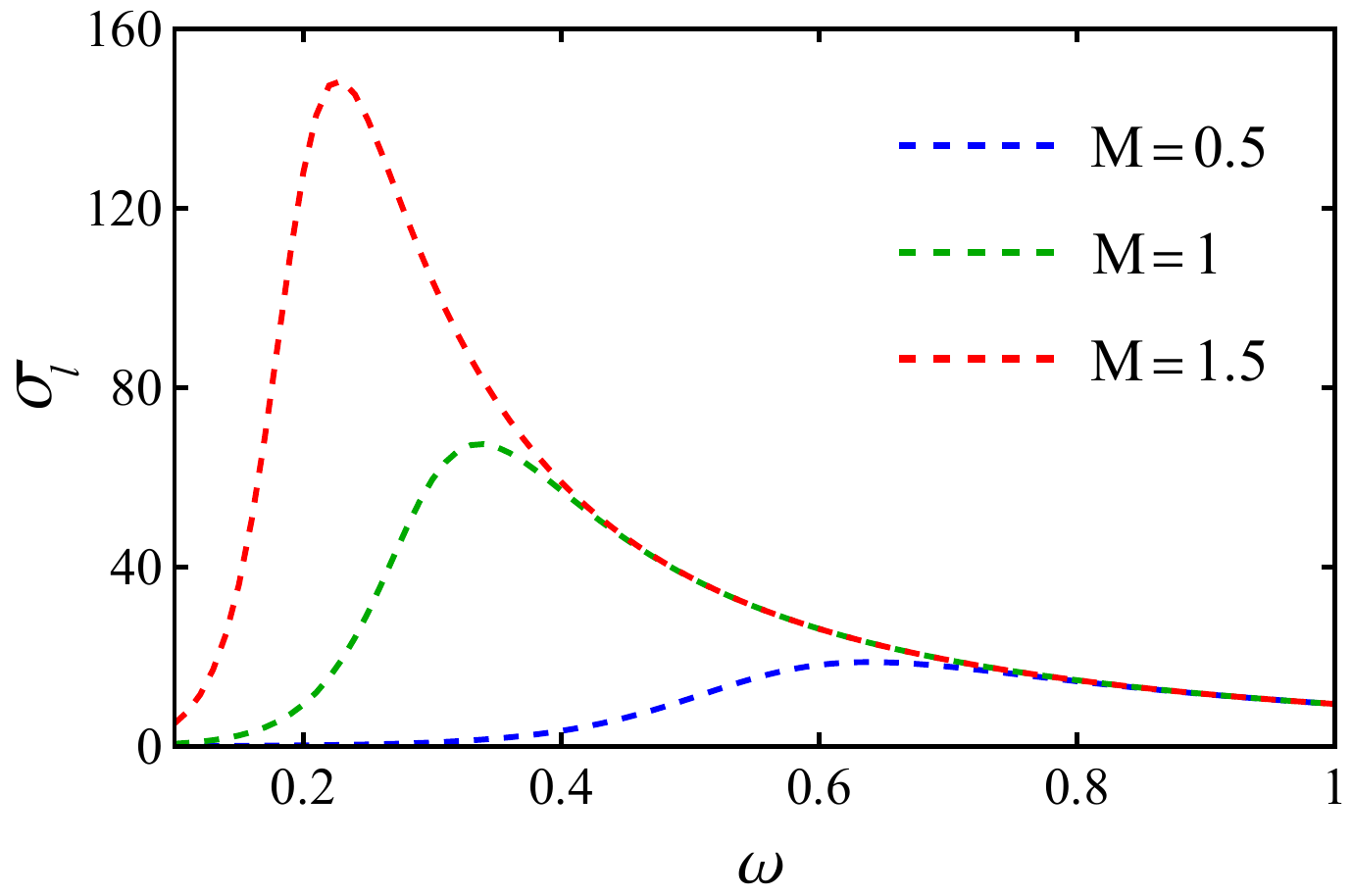}
	\caption{Partial absorption cross section for $l = 1$, $\Theta^2=0.4$ with different mass $M = 0.05,0.1,0.15,0.5,1$}
	\label{fig:cross3}
\end{figure}

In Fig. \ref{fig:cross1}, we can see the partial absorption cross sections plotted against frequency for both $l=1$ and $l=1$ to $l=3$, with varying values of the NC parameter. The figure shows that larger values of the NC parameter correspond to higher partial absorption cross sections. This trend can be explained by the fact that as the height of the effective potential decreases with increasing values of the NC parameter, the absorption of the scalar field increases for a given frequency $\tilde{\omega}$. Fig. \ref{fig:cross2} further illustrates this behavior by displaying how the absorption cross section increases with increasing values of $\Theta$. In Fig. \ref{fig:cross3}, we also display the partial absorption cross section of different values of $M$.
\\

\section{GEODESICS IN THE NON--COMMUTATIVE MODEL }\label{sec6}

Geodesics are of paramount importance in physics, as they reveal the curvature of space-time and the behavior of particles under gravitational influences. Such a study in NC scenarios has emerged as a promising field of research, as it explores the implications of quantum effects on the properties of spacetime. Understanding geodesics in NC scenarios can help us elucidate the behavior of particles and fields at extremely small scales, where quantum effects become dominant. Moreover, it is crucial to comprehend the geodesic configuration of NC BHs to interpret and scrutinize astrophysical observations related to these entities, such as the features of accretion disks and shadows.

In this direction, we make this section in order to provide such an investigation. The geodesic equation reads,
\ie
\frac{\mathrm{d}^{2}x^{\mu}}{\mathrm{d}s^{2}} + \Gamma\indices{^\mu_\alpha_\beta}\frac{\mathrm{d}x^{\alpha}}{\mathrm{d}s}\frac{\mathrm{d}x^{\beta}}{\mathrm{d}s} = 0. \label{geodesicfull}
\fe
Our fundamental aim is to investigate the influence of non-commutativity on the trajectories of massless and massive particles. This requires solving a set of lengthy partial differential equations that arise from Eq. (\ref{geodesicfull}). In particular, above expression yields four coupled partial differential equations that must be solved

\ie
t''(s) = \frac{\left(3 \Theta ^2+64 M^2\right) r'(s) t'(s)}{r(s) \left(3 \Theta ^2+64 M^2-32 M r(s)\right)},
\fe
\ie
\begin{split}
r''(s) = & \frac{\left(3 \Theta ^2+64 M^2-32 M r(s)\right) \left(\left(3 \Theta ^2+64 M^2\right) t'^2(s) - 64 M r^3(s) \left(\theta'^2+\sin ^2(\theta ) \varphi'^2(s)\right)\right)}{2048 M^2 r(s)^3} \\
& +\frac{\left(3 \Theta ^2+64 M^2\right) r'^2(s)}{2 r(s) \left(-3 \Theta ^2-64 M^2+32 M r(s)\right)},
\end{split}
\fe
\ie
\theta''(s) = \sin (\theta ) \cos (\theta ) \left(\varphi'(s)\right)^2-\frac{2 \theta' r'(s)}{r(s)},
\fe
\ie
\varphi''(s) = -\frac{2 \varphi'(s) \left(r'(s)+r(s) \theta' \cot (\theta )\right)}{r(s)},
\fe
where, $s$ is an arbitrary parameter, and the prime symbol ``$\prime$'' denotes differentiation with respect to it (i.e., $\mathrm{d}/\mathrm{d}s$).

It is evident that the above equations do not have analytical solutions, and therefore a numerical analysis is necessary to examine the geodesic trajectories of massive particles. In order to achieve this, we select certain fixed parameters, such as $\phi$ and $\theta$, and solve them. Fig. \ref{fig:trajectory} depicts the deflection of light for various values of $\Theta$. Notably, the NC parameter has a significant impact on the path of light, causing a ``contraction'' to it as $\Theta$ increases. Furthermore, we observe that the event horizon expands as $\Theta$ increases. Recent research has also explored the divergent reflections of BHs resulting from non--commutativity \cite{snepppen2021divergent}.

\begin{figure}\label{fig:trajectory}
	\centering
	\includegraphics[scale=0.55]{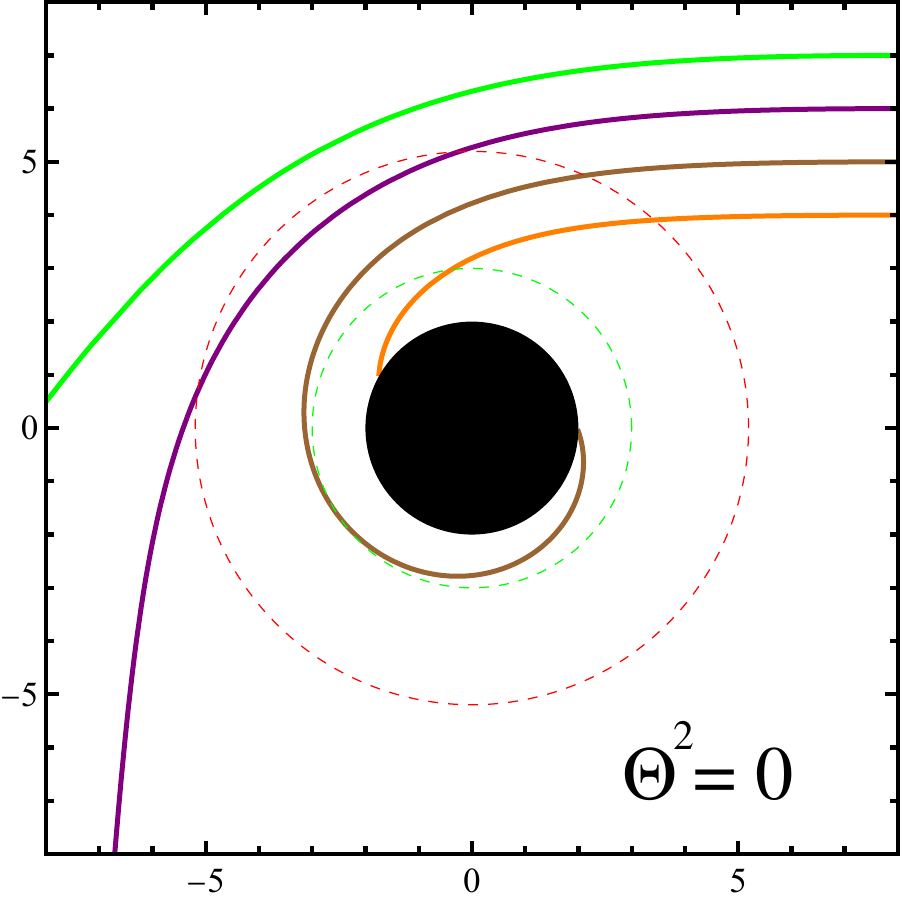}
	\includegraphics[scale=0.55]{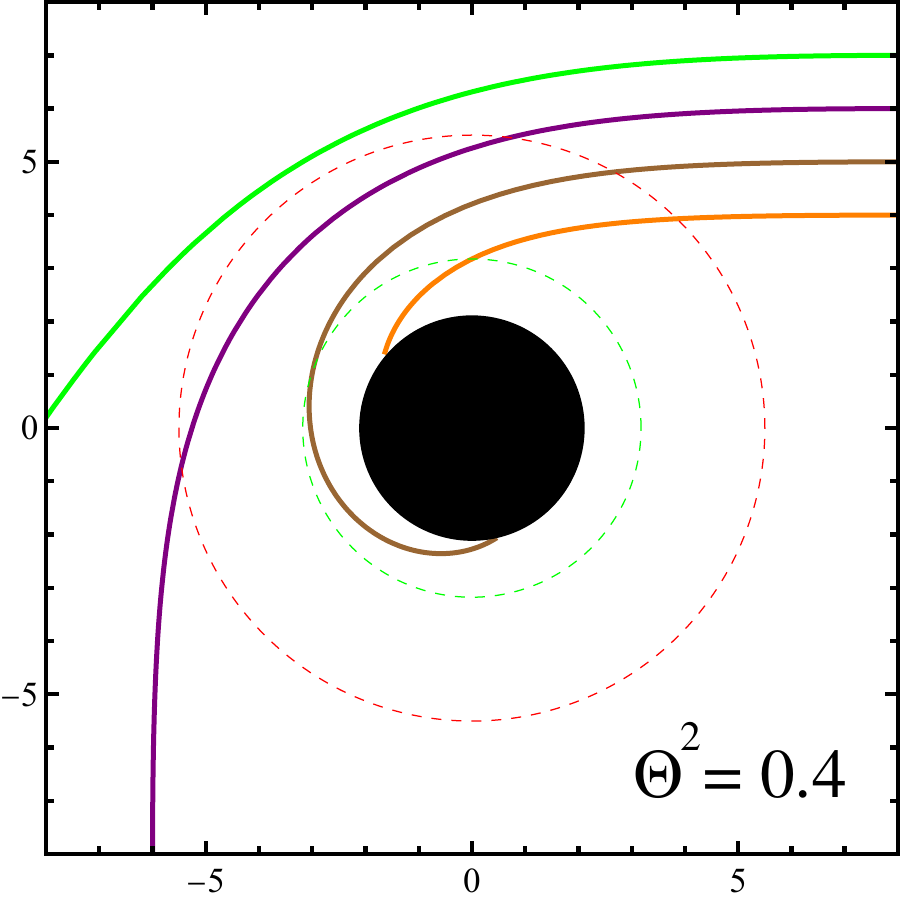}
	\includegraphics[scale=0.55]{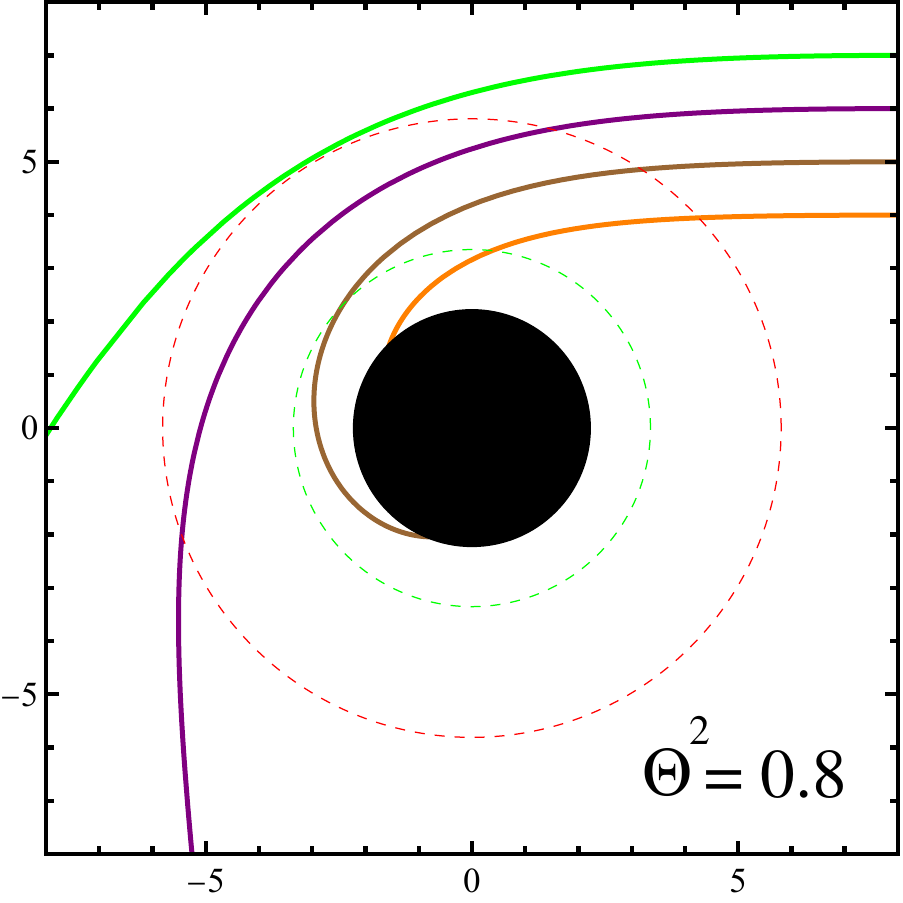}
	\includegraphics[scale=0.55]{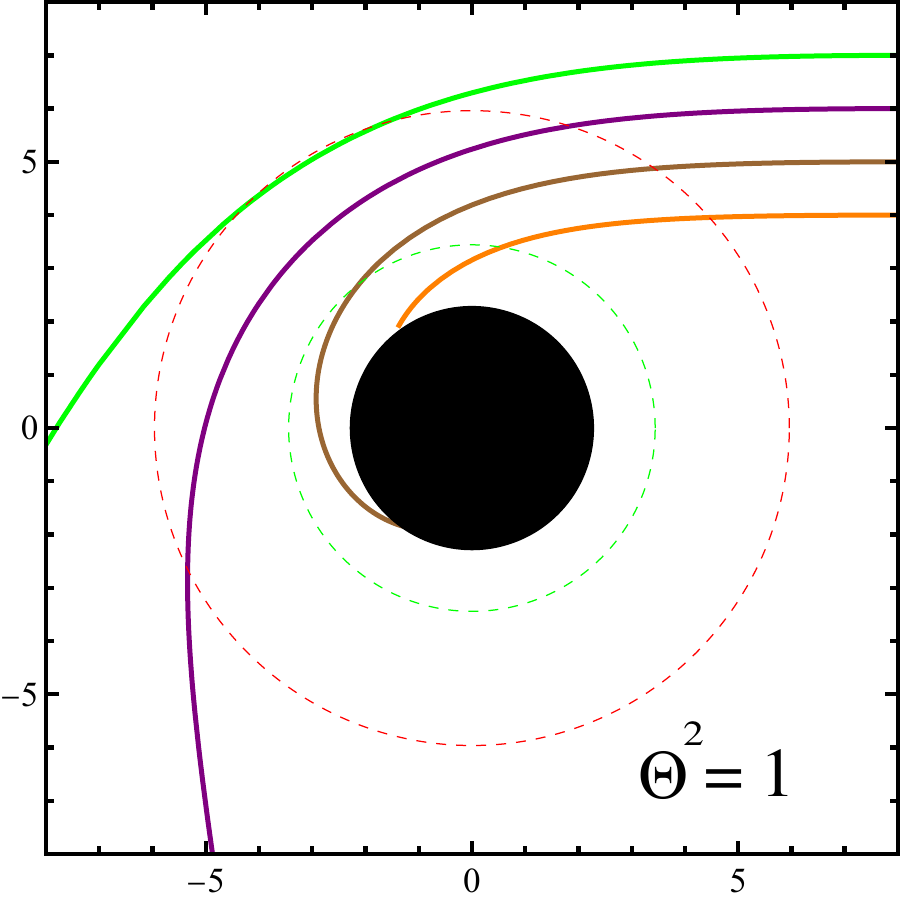}
	\caption{Four light trajectory are shown with different initial condition of impact factor $\xi=4,5,6,7$ for the same mass $M = 1$ and $ \Theta^2 = 0 , 0.4, 0.8$ and $1$. The dotted green lines represent the photon sphere while the red dotted correspond to the shadows of the BH. } \label{fig:trajectory}
\end{figure}

\begin{figure}
    \centering
    \includegraphics[scale=0.4]{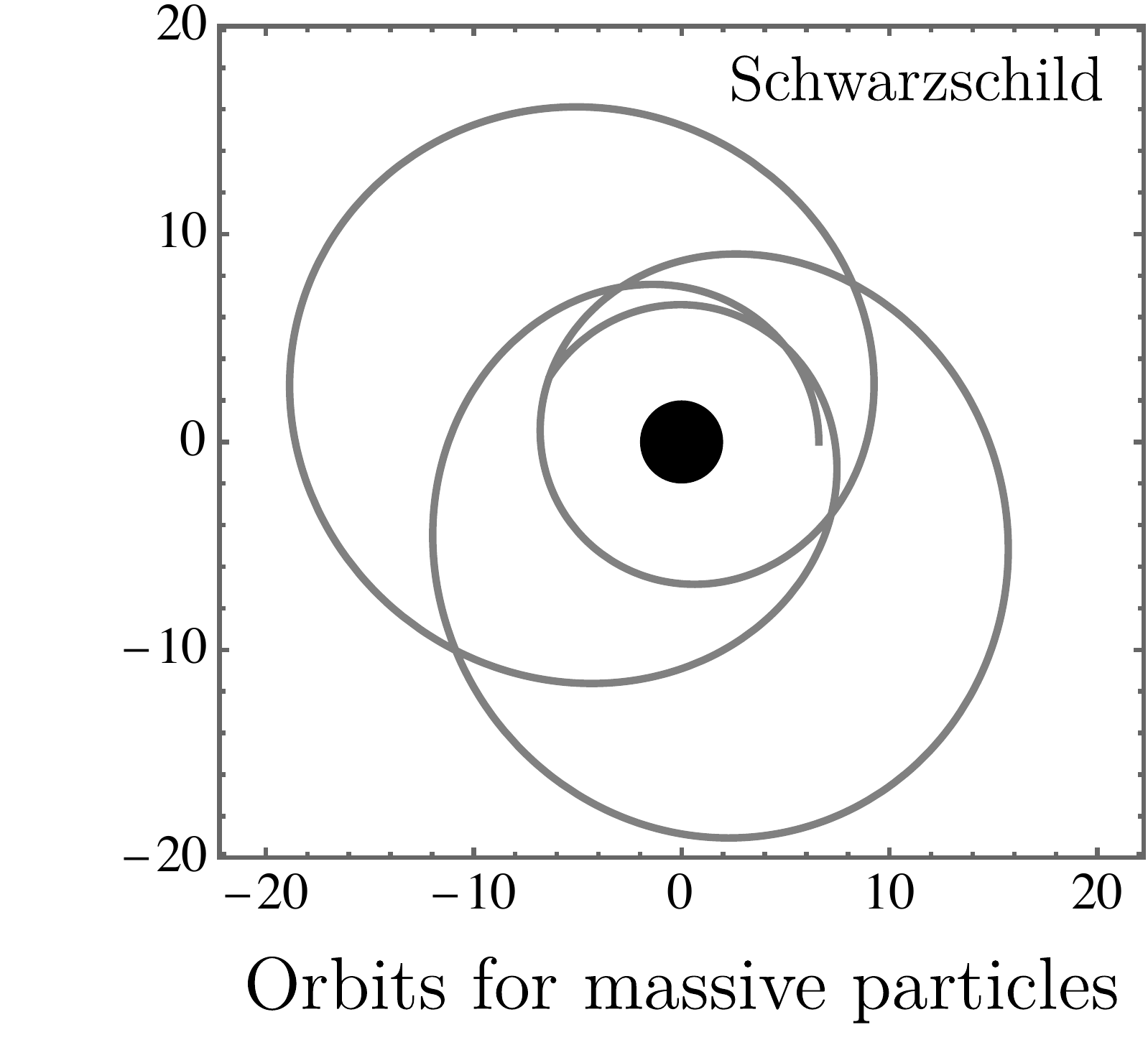}
    \includegraphics[scale=0.4]{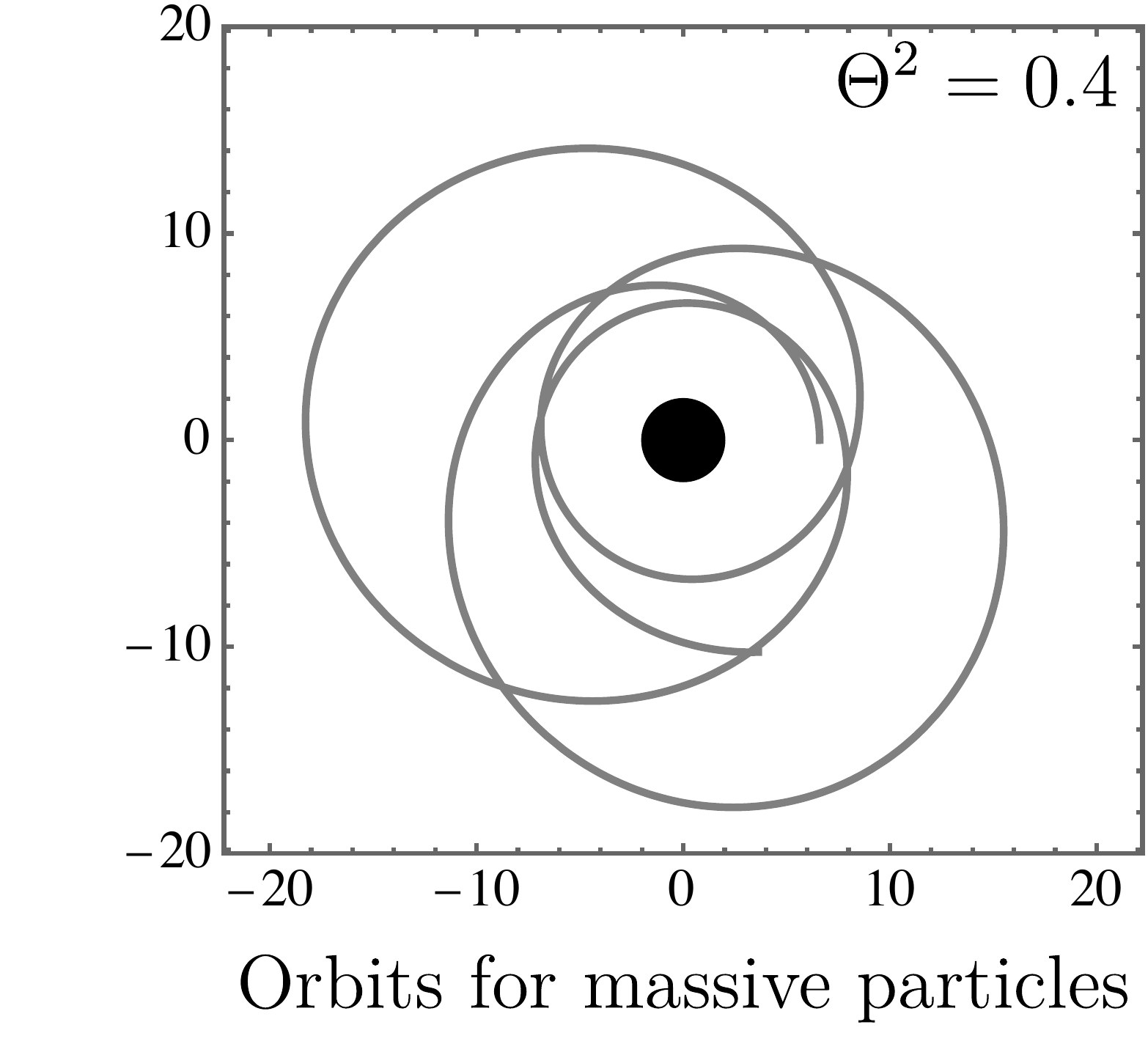}
    \includegraphics[scale=0.4]{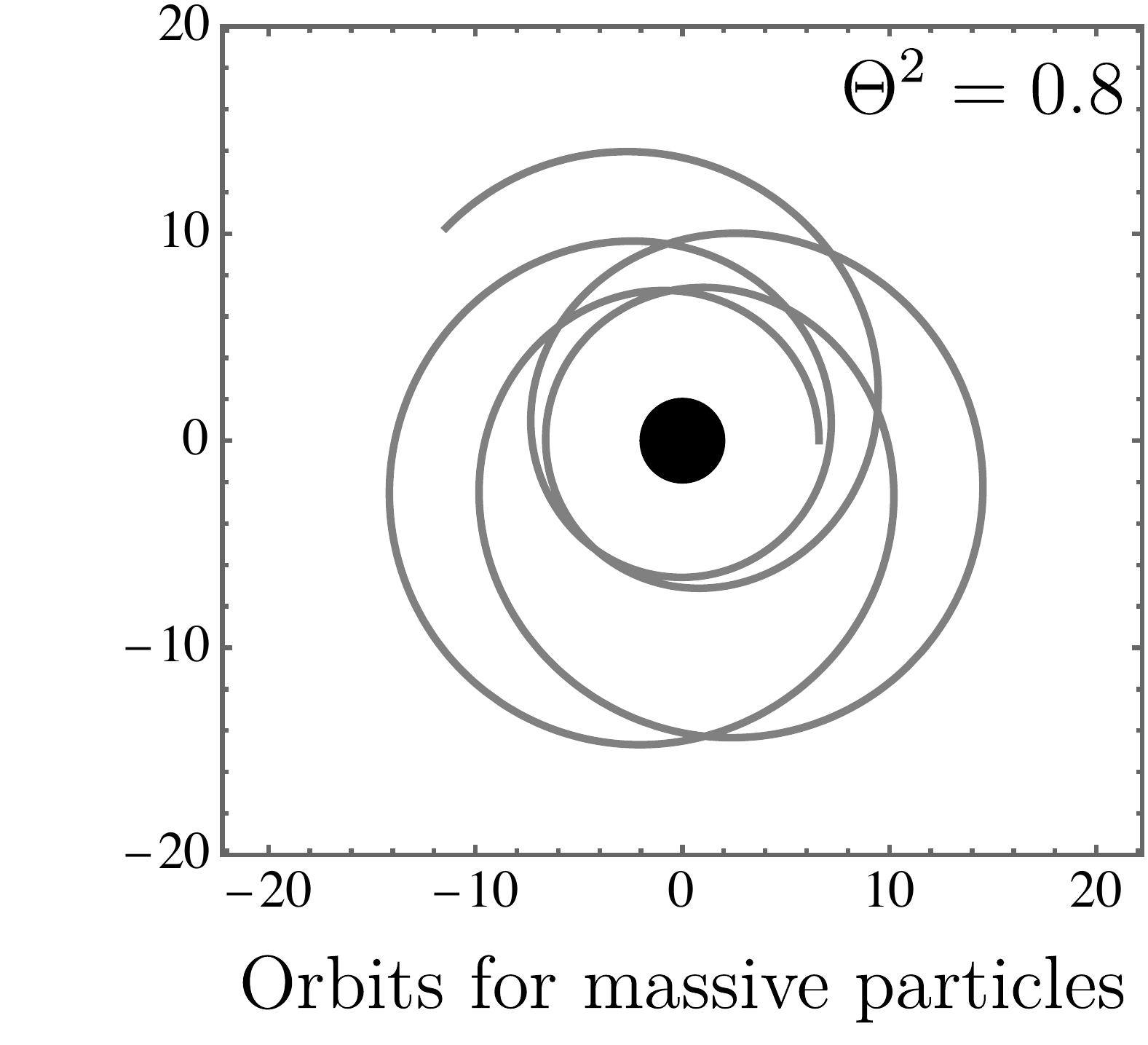}
    \includegraphics[scale=0.4]{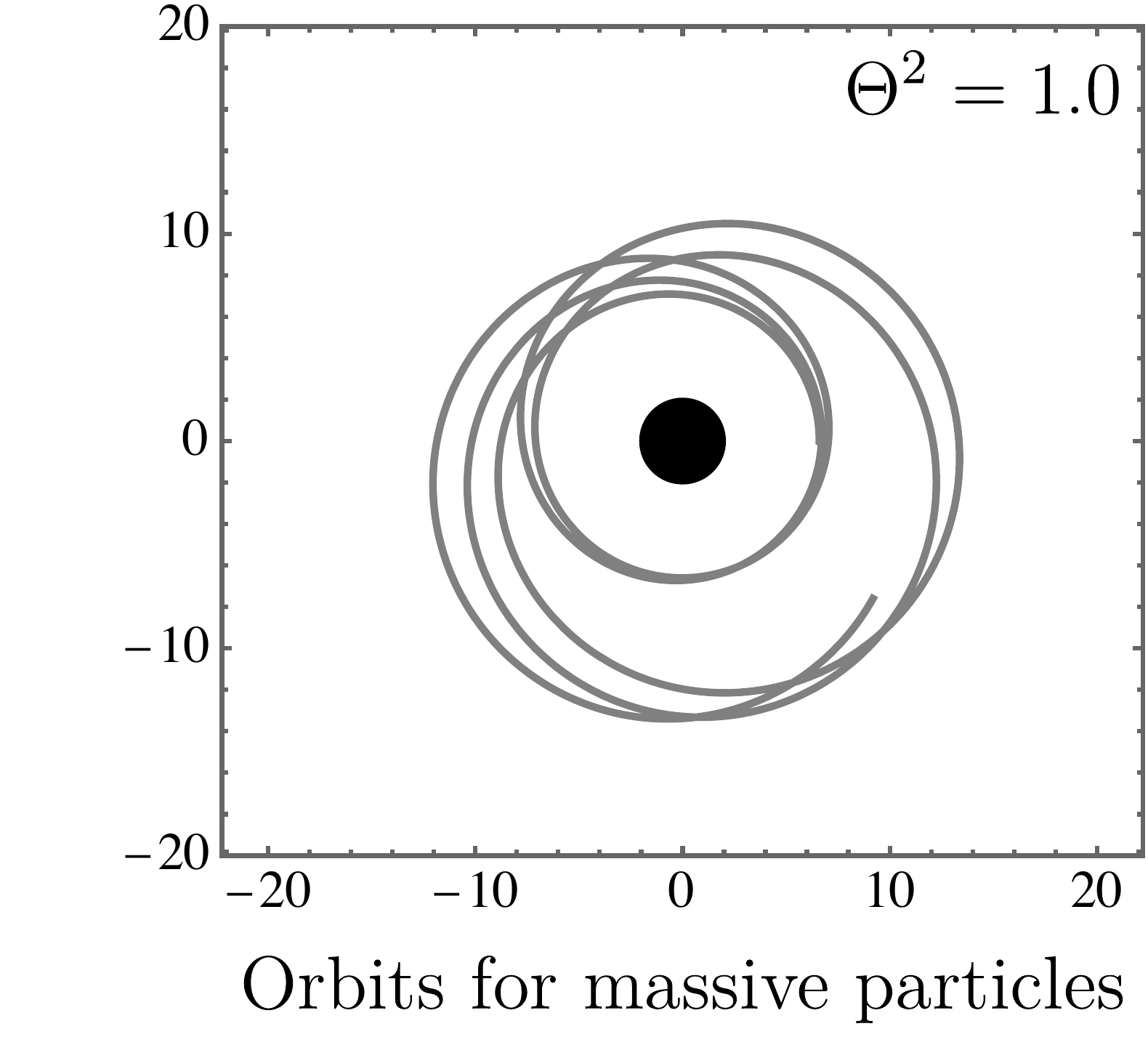}
    \caption{The trajectories of massive particles are exhibited for $r = 6.6$, $\theta=\varphi = \pi/2$.}
    \label{deflectionoofmassiveparticles}
\end{figure}

In contrast, Fig. \ref{deflectionoofmassiveparticles} illustrates the behavior of trajectories for massive particles. As expected, these particles also experience significant modifications compared to massless particles. For different values of $\Theta$, we observe a ``squeezing'' behavior and a larger event horizon, similar to the massless case. To this case, the ``squeezing'' effect is more evident in comparison with the light case. It is noteworthy that all system configurations are compared to the Schwarzschild case. It is worth mentioning that the advance of Mercury orbit has also been calculated within the context of NC scenario, with an estimate of $\Theta^{\mu\nu}$ based on observational data \cite{f5}. This emphasizes the importance of studying the effects of non--commutativity in gravitational systems.

\subsection{Critical orbits in the non--commutative model}

The critical orbits are essential to comprehend the dynamics of particles and the behavior of light rays in the vicinity of BHs with NC structures. These orbits offer valuable insights into the properties of spacetime under quantum effects.

To better understand the consequences of non--commutativity on the photon sphere (critical orbit) of our BH, we shall use the Lagrangian method to develop the calculations of null--geodesics. This method provides a more accessible approach for readers to comprehend the calculations if compared to the geodesic equation. By examining the effects of $\Theta$ on the photon sphere, we can gain an additional information of the gravitational effects of non--commutativity and their potential implications for observational astronomy. To begin our investigation, we express the Lagrangian method as:

\begin{equation}
\mathcal{L} = \frac{1}{2} g_{\mu\nu}\Dot{x}^{\mu}\Dot{x}^{\nu},
\end{equation}

where for a fixed angle of $\theta=\pi/2$, the above expression reduces to:
\ie
g_{00}^{-1} E^{2} + g_{11}^{-1} \Dot{r}^{2} + g_{33}L^{2} = 0,
\fe
where $E$ is the energy and $L$ the angular momentum. Above expression can also be rewritten as
\ie
\Dot{r}^{2} = E^{2} - \left( 1 - \frac{2M_{\Theta}}{r} \right)\left(  \frac{L^{2}}{r^{2}} \right),
\fe
with $\overset{\nsim}{V} \equiv \left( 1 - \frac{2M_{\Theta}}{r} \right)\left(  \frac{L^{2}}{r^{2}} \right)$ is the effective potential. With it, we can straightforwardly address to the critical radius as $r_{c} = 3 \left( M + \frac{3\Theta^{2}}{64M} \right)$. Table \ref{criticalradius} summarizes the impact of non--commutativity on the critical orbit, aiming to aid the understanding of $r_{c}$. One notable fact gives rise to: the photon sphere turns out to be larger as the NC parameter $\Theta$ increases. This finding aligns with the outcomes illustrated in our manuscript regarding light deflection (refer to Fig. \ref{fig:trajectory}).

 
 
 
 
 
 
 
 
 

\begin{table}
	\centering
	\begin{tabular}{|l|l|l|l|l|l|l|l|l|l|l|l|}
		\hline 
		$\Theta^{2}$      & 0.0 & 0.1 & 0.2 & 0.3 & 0.4 & 0.5 & 0.6 & 0.7 & 0.8 & 0.9 & 1.0 \\
		\hline 
		 $r_{c}$ & 3.00000 & 3.01406 & 3.02812 & 3.04219  & 3.05625 & 3.07031 & 3.08437 & 3.09844 & 3.11250 & 3.12656 & 3.14063 \\
		\hline 
	\end{tabular}
	\caption{The value of the critical orbit $r_{c}$ changes in response to variations in the NC parameter $\Theta$, considering $M=1$.}
	\label{criticalradius}
\end{table}

\section{SHADOW RADIUS IN NON--COMMUTATIVE MODEL}\label{sec7}

To investigate the null geodesics equation in the spacetime of deformed Schwarzschild spacetime and study the effect of the NC parameter on photon evolution, we begin by considering the spherically symmetric spacetime metric given by Eq. (\ref{metric}) and apply the Hamilton--Jacobi action.
\begin{equation}\label{action1}
\frac{{\partial S}}{{\partial \tau }} =  - \frac{1}{2}{g^{\mu \nu }}\frac{{\partial S}}{{\partial {\tau ^\mu }}}\frac{{\partial S}}{{\partial {\tau ^\nu }}}
\end{equation}
where $S$ is the Jacobi action and $\tau$ is the arbitrary affine parameter. The variables can be separated by
\begin{equation}\label{action2}
S = \frac{1}{2}{m^2}\tau  - Et + L\phi  + {S_r}(r) + {S_\theta }(\theta ).
\end{equation}
We can express $S_r(r)$ and $S_{\theta}(\theta)$ as functions of $r$ and $\theta$, respectively. Since we are considering a photon with zero mass, $m$ is zero. The constants of motion, energy $E$ and angular momentum $L$, are conserved quantities along the photon's path. By using Eqs. (\ref{action1}) and (\ref{action2}), we can obtain the null geodesic equations as follows:
\begin{equation}\label{time}
\frac{{\mathrm{d}t}}{{\mathrm{d}\tau }} = \frac{E}{{f_{\Theta}(r)}},
\end{equation}
\begin{equation}\label{position}
\frac{{\mathrm{d}r}}{{\mathrm{d}\tau }} = \frac{{\sqrt {\mathcal{R}(r)} }}{{{r^2}}},
\end{equation}
\begin{equation}\label{thetadot}
\frac{{\mathrm{d}\theta }}{{\mathrm{d}\tau }} =\pm \frac{{\sqrt {\mathcal{Q} (r)} }}{{{r^2}}},
\end{equation}
\begin{equation}\label{phidot}
\frac{{\mathrm{d}\varphi }}{{\mathrm{d}\tau }} = \frac{{L\,{{\csc }^2}\theta }}{{{r^2}}},
\end{equation}
where $\mathcal{R}(r)$ and $\mathcal{Q}(\theta)$ are defined as 
\begin{equation}
\mathcal{R}(r) = {E^2}{r^4} - (\mathcal{K} + {L^2}){r^2}f_{\Theta}(r)
\end{equation}
\begin{equation}
\mathcal{Q} (\theta ) = \mathcal{K} - {L^2}{\cot ^2}\theta,
\end{equation}
and $\mathcal{K}$ is the Carter constant \cite{carter1968global}. The plus and minus sign in Eq. (\ref{thetadot}) indicate a motion of the photon in the outgoing and ingoing radial direction. In this work, we set the angle $\theta=\pi/2$ and consider only the equatorial plane without loss of generality. Now, we focus on the radial equation by introducing the effective potential
\begin{equation}\label{veff2}
{\left(\frac{{\mathrm{d}r}}{{\mathrm{d}\tau }}\right)^2} + {\mathcal{V}_{eff}}(r) = 0.
\end{equation}
Here, the effective potential is defined 
\begin{equation}\label{veffr}
{\mathcal{V}_{eff}}(r) = {(L^2+\mathcal{K}) }\frac{f_\Theta(r)}{{{r^2}}} - {E^2}
\end{equation}
and we introduce  two impact parameter
\begin{equation}\label{par}
\xi  = \frac{L}{E}
\text{ and }
\eta  = \frac{\mathcal{K}}{{E^2}}.
\end{equation}
The critical radius of a photon can be calculated by applying the unstable condition on effective potential as 
\begin{equation}
{\mathcal{V}_{eff}} = \frac{{\mathrm{d}{\mathcal{V}_{eff}}}}{{\mathrm{d}r}} = 0,
\end{equation}
which leads
\begin{equation}\label{rps}
2 - \frac{{rf_{\Theta}'(r)}}{{f_{\Theta}(r)}}{\arrowvert_{r = {r_{c}}}} = 0.
\end{equation}
Also, by using both Eq. (\ref{veffr}) and Eq. (\ref{par}) in above expression, we obtain 
\begin{equation}
{\xi ^2} + \eta  = \frac{{r_{c}^2}}{{f_{\Theta}({r_{c}})}}.
\end{equation}

Let us now calculate the shadow radius using celestial coordinates $\alpha$ and $\beta$ \cite{singh2018shadow}, which are related to the constants of motion as $\alpha=-\xi$ and $\beta=\pm\sqrt{\eta}$. The shadow radius can be expressed as
\begin{equation}
{R_{sh}} = \sqrt {{\alpha ^2} + {\beta ^2}} = \frac{{{r_{c}}}}{{\sqrt {f_{\Theta}({r_{c}})} }},
\end{equation}
where ${r_{c}}$ is the radius of the photon sphere obtained in Eq. (\ref{rps}). In this manner, the equation for the shadow radius in the NC spacetime reads
\begin{equation}
{R_{sh}} = 3\sqrt 3 M\left(1 + \frac{{3\pi {\Theta ^2}}}{{64 M^2}}\right).
\end{equation}

\begin{figure}[!]
	\centering
	\includegraphics[width=80mm]{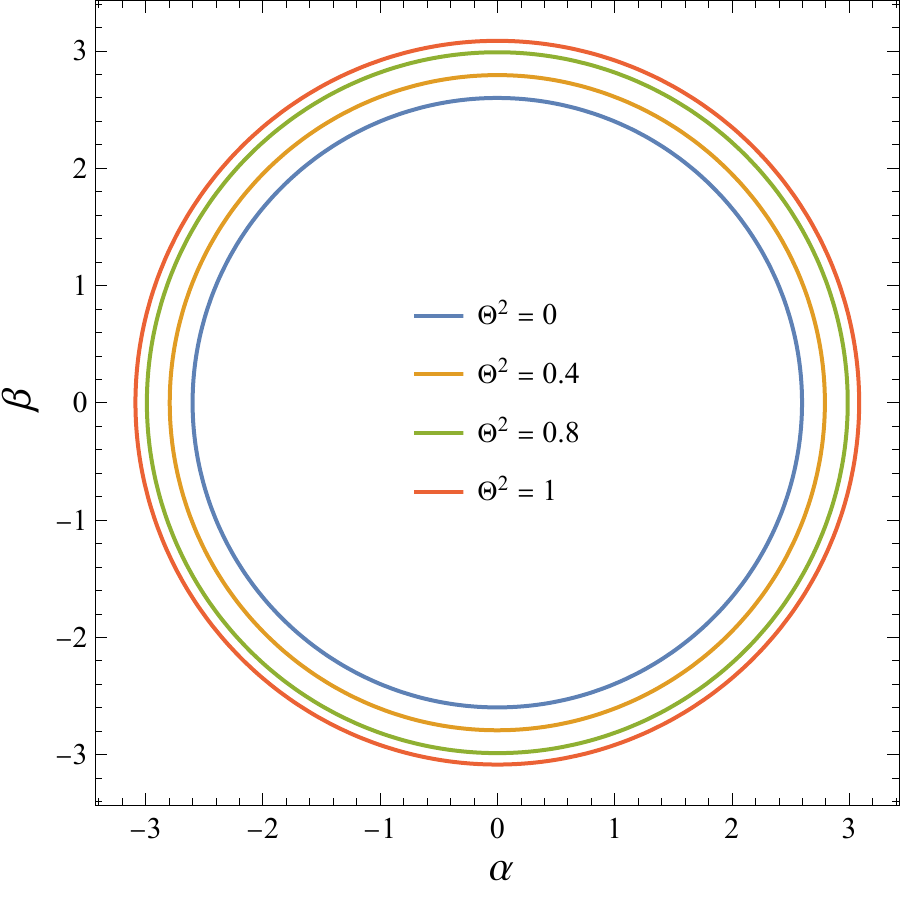}
	\hfill
	\includegraphics[width=80mm]{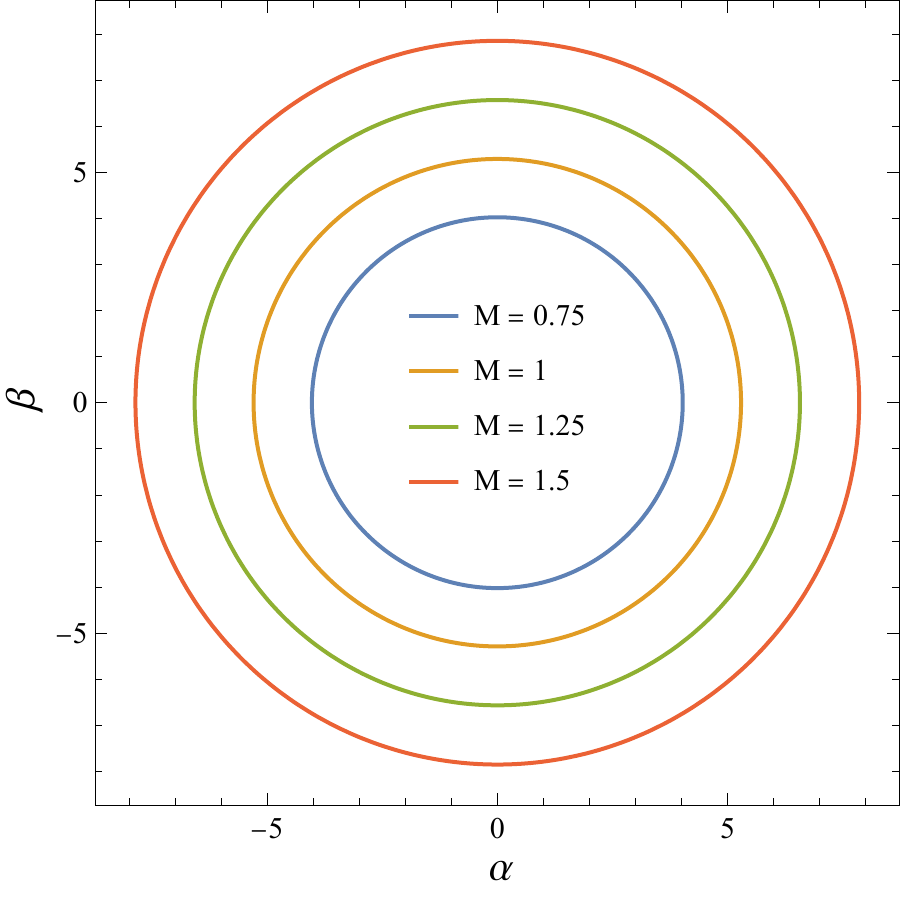}
	\caption{(left) Shadow of the BH with mass $M=1$ for non--commutativity parameter values of $\Theta^2=0, 0.4, 0.8, 1$. (right) Shadow radius for a BH with $\Theta^2=0.4$ and various masses: $M=0.15,0.25,0.35,0.45,0.55$.}
	\label{fig:shadow}
\end{figure}

In Fig. \ref{fig:shadow}, we present an analysis of the shadow boundaries of a NC BH for a range of $\Theta$ values on the left--hand side. Notably, the shadow radius exhibits an increase as the $\Theta$ parameter rises. On the right-hand side, we explore the impact of the BH's mass on the same NC parameter. As evidenced by the figure, the shadow radius undergoes an increase when the original mass $M$ transitions from $0.75$ to $1.5$. As the remnant mass at the fixed condition of NC parameter of $\Theta^2=0.4$ is $M_{rem} \sim 0.137$, the mass is considered to be changed beyond the remnant boundary.

It is pertinent to note that the deformed NC mass $M_{\Theta}$ defined in Eq. (\ref{mass}) attains a minimum value. The shadow radius will exhibit a decrease until it attains this minimum value when the original mass of the BH is reduced. As we surpass this point, an increase in the original mass of the BH leads to a rise in the deformed mass $M_{\Theta}$, subsequently resulting in an increase in the shadow radius.

\begin{figure}[h]
	\centering
	\includegraphics[width=100mm]{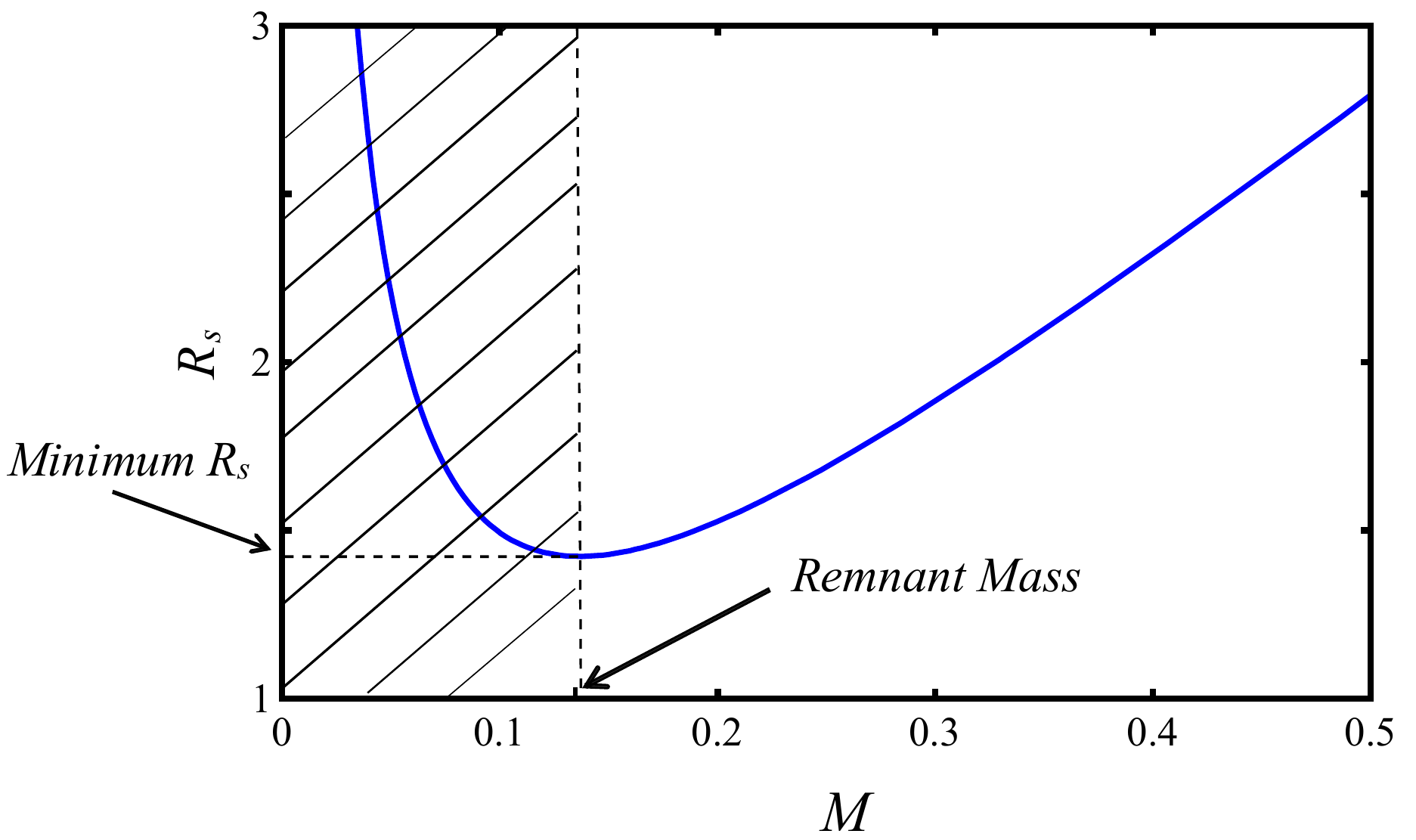}
	\caption{Shadow radius is represented versus $M$ for $\Theta^2=0.4$, the remnant mass and its related shadow radius is shown.}
	\label{fig:Ref}
\end{figure}

To enhance the visual representation, following the approach in \cite{jusufi2020quasinormal}, we present a plot in Fig. \ref{fig:Ref} illustrating the relationship between the shadow radius $R_s$ and the mass $M$, while keeping $\Theta$ fixed. Notably, the inclusion of the NC effect introduces a minimum remnant mass $M$, which subsequently leads to a minimum shadow radius. It is important to note that the dashed area depicted in Fig. \ref{fig:shadow} is considered unphysical due to these NC effects.

\begin{figure}[h]
	\centering
	\includegraphics[width=90mm]{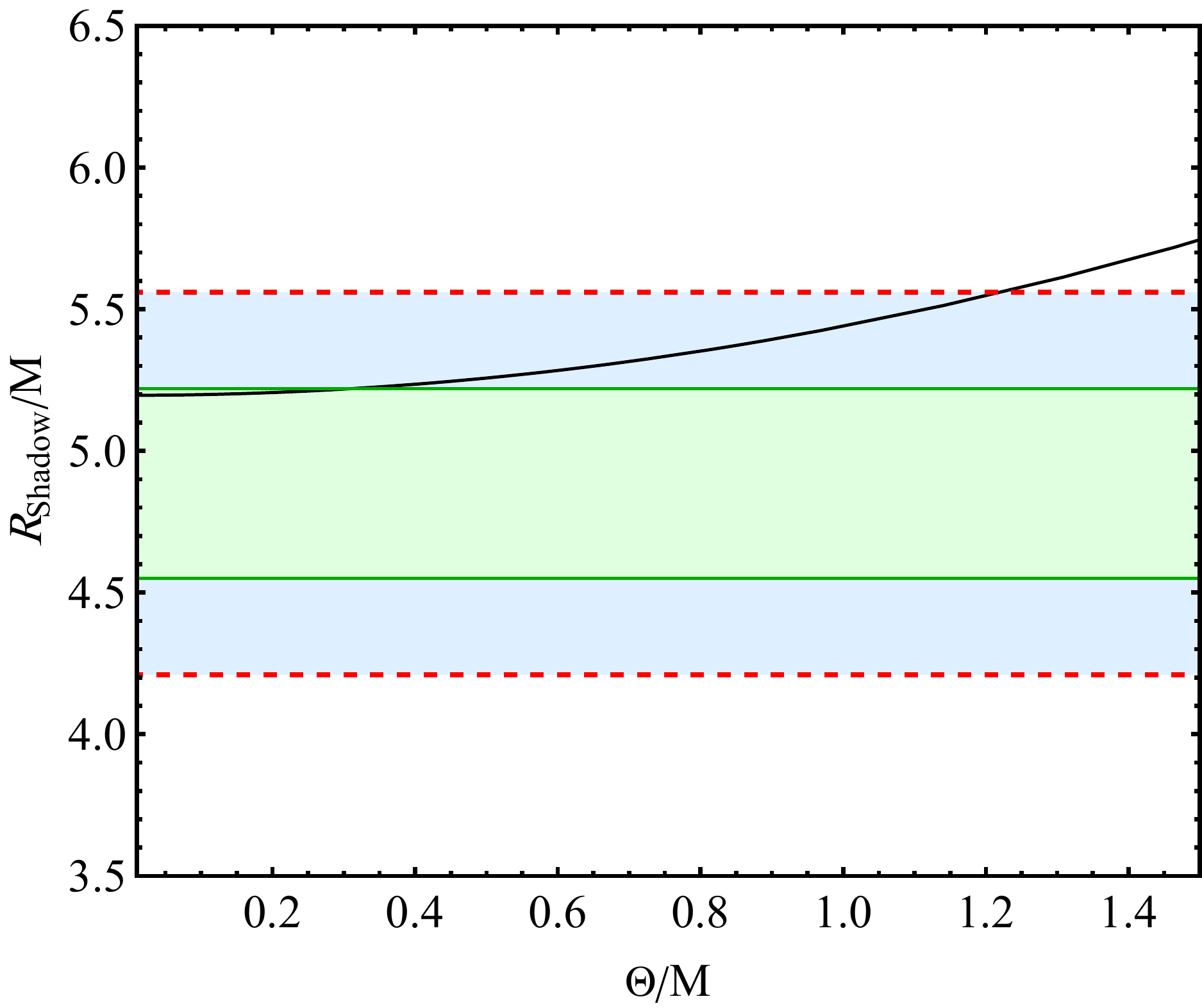}
	\caption{ Shadow radius in units of mass is plotted versus $\Theta$ in units of mass, the experimental constraints are represented with green and blue according to Eq. (\ref{const1}) and Eq. (\ref{const2}), respectively.}
	\label{fig:constraint}
\end{figure}

According to the EHT horizonscale of $Sgr A^*$,  the Keck and VLTI mass--to--distance ration priors for $Sgr A^*$ are averaged and considering two standard deviation two constraints have been obtained for shadow radius. \cite{vagnozzi2022horizon,akiyama2022first}.
\begin{equation}\label{const1}
4.55 < \frac{R_{sh}}{M} < 5.22,
\end{equation}
and
\begin{equation}\label{const2}
4.21 < \frac{R_{sh}}{M} < 5.56.
\end{equation}
We have examined the proper constraint for NC parameter upon the experimental constraints. The shadow radius in units of mass ($M$) is plotted with respect to the NC parameters in units of $M$ in Fig.\ref{fig:constraint}. The blue and green area are consistent to the experimental limits in Eqs. (\ref{const1},\ref{const2}).
The plot cuts the experimental constraint lines in two points which make an upper limit on NC parameter. Therefore, we have $\Theta<0.312$ and $\Theta<1.222$, based on first and second experimental data, respectively.

\section{DEFLECTION ANGLE IN NON--COMMUTATIVE MODEL}\label{sec8}

The deflection of light as it propagates through curved spacetime represents a pivotal and captivating phenomenon, serving as an indispensable tool in the realm of scientific inquiry. By elegantly conforming to the contours of spacetime, light unveils a mesmerizing interplay that bestows invaluable insights into the intricate physics underlying gravitational sources \cite{70,71,72,73,74,75}. Through meticulous observation and rigorous analysis, this phenomenon not only unlocks profound discoveries but also engenders a deeper comprehension of the cosmos, fostering the expansion of our knowledge and illuminating the fundamental workings of the universe.

Thereby, to quantify this deflection angle, a well--established formula \cite{76,jusufi2020quasinormal} is employed
	\begin{equation}\label{phi}
	\hat \alpha ({r_{\min }}) = 2\int_{{r_{\min }}}^\infty  {\frac{\mathrm{d}r}{{r\sqrt {{{(\frac{r}{{{r_{\min }}}})}^2}{f_\Theta }({r_{\min }}) - {f_\Theta }(r)} }} - \pi } 
	\end{equation}
	where $r_{min}$ is the distance of closest approach of light ray to the BH which is linked to the impact parameter as
	\begin{equation}
	\xi = \frac{{{r_{\min }}}}{{\sqrt {1 - \frac{{2M}}{{{r_{\min }}}}} }}.
	\end{equation}
 
In order to assess the influence of the NC parameter on the deflection angle, we proceed by substituting the metric derived from Equation (\ref{metric}) into Equation (\ref{phi}). The obtained results are summarized in Table \ref{tab:deflection}, elucidating the deflection angle for a NC Schwarzschild BH with a constant impact parameter and variable values of $\Theta$. Moreover, to further investigate this relationship, Table \ref{tab:deflectionmass} presents the corresponding deflection angles for distinct values of $\Theta$, while maintaining the mass fixed at $M=1$. 

Furthermore, it is noteworthy to observe that the deflection angle exhibits a direct correlation with both the mass and the parameter $\Theta$. Specifically, as the values of mass and $\Theta$ increase, there is a corresponding increase in the magnitude of the deflection angle. This finding underscores the influence of these factors on the bending of light in the presence of a NC Schwarzschild BH, emphasizing the significant role played by mass and parameter $\Theta$.
	
\begin{table}[!ht]
		\centering

			\caption{\label{tab:deflection}Minimum radius and deflection angel for $M=1$ and $\Theta^2=0,0.4,0.8$ and $1$.}
		
		\begin{tabular}{|l|l|l|l|l|}
			\hline\hline
			$\Theta^2$ & 0 & 0.4 & 0.8 & 1 \\ \hline\hline
			$r_{min}$ & 8.78885
			 & 8.76022
			  & 8.73125
			   & 8.71664
			    \\ \hline
			deflection angel & 0.590396
			 & 0.607591
			  & 0.625183
			   & 0.634132
			    \\ \hline
		\end{tabular}
	\end{table}
	\begin{table}[!ht]
		\centering

			\caption{\label{tab:deflectionmass} Minimum radius and deflection angel for $\Theta^2=0.4$ and $M=0.75,1,1.25$  and $1.5$.}
		
		\begin{tabular}{|l|l|l|l|l|}
			\hline\hline
			$M$ & 0.75 & 1 & 1.25 & 1.5\\ \hline\hline
			$r_{min}$ & 9.10961 & 8.76022 & 8.34832 & 7.83537 \\ \hline
			deflection angel & 0.409805 & 0.607591 & 0.878165 & 1.29283 \\ \hline
		\end{tabular}
	\end{table}

\section{Conclusion}\label{conclusion}

Our study aimed to comprehensively explore various fundamental aspects of a non--commutative (NC) theory, with a specific emphasis on the implications of mass deformation. In pursuit of this objective, we conducted a thorough investigation into the system's thermodynamic properties, comparing our findings with the latest research documented in the literature \cite{araujo2023thermodynamics}. Additionally, we delved into the analysis of quasinormal modes associated with massless scalar perturbations, employing two distinct methodologies: the WKB approximation and the Pöschl--Teller fitting technique.

Our examination of the Klein--Gordon equation allowed us to determine the effective potential, which we successfully matched with the Pöschl--Teller potential, serving as our chosen effective potential. This approach enabled us to analytically compute the quasinormal modes. Notably, our analysis unveiled a noteworthy characteristic: the real part of the quasinormal modes remained independent of the quantum number $n$ contrasting with numerical results suggesting an $n$--dependent behavior. Remarkably, our adoption of the Pöschl--Teller potential aligned with numerical findings, demonstrating this $n$--dependent aspect.

Furthermore, we observed that increasing the NC parameter led to a decrease in both the real and imaginary parts of the quasinormal modes, indicating reduced stability and decay rates with higher NC values. Additionally, the NC parameter increment resulted in an overall rise in the total absorption cross--section across all $\omega$ values.

Our investigation extended to the analysis of geodesics for both massless and massive particles, with a specific focus on the influence exerted by the NC parameter $\Theta$. Our findings emphasized the substantial impact of $\Theta$ on the trajectories of light and event horizons. Specifically, we noted that an increasing NC parameter corresponded to heightened light absorption by the black hole, resulting in an expansion of the photonic radius and the size of the shadow.

Additionally, we harnessed observational data from the EHT related to the Schwarzschild NC black hole shadow of $Sgr A^*$ to establish upper limits on the NC parameter. Our analysis, incorporating EHT constraints on $Sgr A^*$'s shadow, compellingly demonstrated that a significant portion of the non--rotating NC black hole parameter space aligned with EHT observations. Our emphasis remained on the non--rotating scenario, as the rotation parameter for $Sgr A^*$ is sufficiently small to induce negligible deviations in the shadow radius \cite{vagnozzi2022horizon}. pauloFinally, our exploration extended to the analysis of the deflection angle within this context.


\section*{Acknowledgements}
The authors would like to thank the anonymous referee for a careful reading of the manuscript and for the remarkable suggestions given to us.
Most of the calculations were performed by using the \textit{Mathematica} software. Particularly, A. A. Araújo Filho is supported by Conselho Nacional de Desenvolvimento Científico e Tecnológico (CNPq) -- [200486/2022--5] and [150891/2023--7].  P. J. Porf\'{\i}rio would like to acknowledge the Brazilian agency CNPq for the financial support, grant No. 307628/2022-1. The author would like to thank R. Konoplya, E. Tollerud, and A. Trounev for providing the \textit{Mathematica} notebook to perform our numerical calculations. Furthermore, we appreciate professor S. Dolan and professor CFB. Macedo and  for their advice on numerical methods.

\section{Data Availability Statement}

Data Availability Statement: No Data associated in the manuscript


\bibliographystyle{ieeetr}
\bibliography{main}

\end{document}